\documentclass[12pt,epsfig]{article} 
\input{epsf.tex}
\usepackage{amssymb,epsfig}
\newcommand{\qed}{QED\begin{math}_3\end{math}\hspace{1ex}}
\newcommand{\One}{1\kern-4.5pt1}
\newcommand{\lapprox}{\raisebox{-0.5ex}{$\ 
\stackrel{\textstyle<}{\textstyle\sim}\ $}}
\newcommand{\gapprox}{\raisebox{-0.5ex}{$\ 
\stackrel{\textstyle>}{\textstyle\sim}\ $}}
\newcommand{\chibchi}{\langle\bar{\chi}\chi\rangle}

\begin{document}
\renewcommand{\topfraction}{.95}
\renewcommand{\textfraction}{.1}

\addtolength{\baselineskip}{0.20\baselineskip}

\rightline{SWAT/06/466}

\hfill September 2006

\vspace{48pt}

\centerline{\Huge Chiral Symmetry Restoration}
\centerline{\Huge in Anisotropic QED$_{3}$ }

\vspace{18pt}

\centerline{\bf
Iorwerth Owain Thomas$^a$ and Simon Hands$^b$}


\vspace{15pt}

\centerline{\sl $^a$Department of Physics, Loughborough University,}
\centerline{\sl Loughborough, Leicestershire LE11 3TU, U.K.}
\smallskip

\centerline{\sl $^b$Department of Physics, Swansea University,}
\centerline{\sl Singleton Park, Swansea SA2 8PP, U.K.}
\vspace{24pt}


\centerline{{\bf Abstract}}

\noindent
{\narrower
We present results from a Monte Carlo simulation of non-compact lattice 
QED in 3 dimensions
in which an explicit anisotropy $\kappa$ between $x$ and $y$ hopping
terms has been introduced into the action. 
Using a parameter set corresponding to broken chiral symmetry in the isotropic
limit $\kappa=1$, we study the chiral condensate on $16^3$, $20^3$, and $24^3$
lattices as $\kappa$ is varied, 
and fit the data to an equation of state which incorporates
anisotropic volume corrections. The value $\kappa_c$ at which chiral symmetry is
apparently restored is strongly volume-dependent, suggesting that the transition
may be a crossover rather than a true phase transition. 
In addition we present results on $16^3$ lattices
for the scalar meson propagator, and for the 
Landau gauge-fixed fermion propagator. The scalar mass approaches the pion mass
at large $\kappa$, consistent with chiral symmetry restoration, but the fermion
remains massive at all values of $\kappa$ studied, suggesting that strong
infra-red fluctuations persist into the chirally symmetric regime.
Implications for models of high-$T_c$ superconductivity based on anisotropic
QED$_3$ are discussed.
}




\vfill
\newpage
\section{Introduction}

\qed$\!\!$, i.e. quantum electrodynamics restricted to two dimensions of space 
and one of time, has recently been the focus of some attention in the condensed
matter community, as various versions of it are examined as candidate effective
models of high-temperature superconductivity in cuprate compounds.

The particular model in which we are interested is that presented in
\cite{Franz:2002qy,Herbut:2002yq}, which is supposed to model
the passage between the
antiferromagnetic SDW (spin-density wave) and superconducting dSC 
(the d denotes the superconducting order parameter has $d$-wave symmetry)
phases at low
temperature $T$ as the doping fraction $x$ is increased.  
QED$_3$ is proposed, as reviewed below in Sec.~\ref{sec:qed3}, as an
effective theory of the low-energy quasiparticle excitations in the
neighbourhood of the 4 nodes in the gap function $\Delta(\vec k)$. Since the
dispersion relation is linear at the nodes, the excitations can be reinterpreted
as various components of a relativistic spinor field $\Psi$ with 4 spin and 
$N_f=2$
flavour degrees of freedom. Interaction via a minimally-coupled
abelian vector gauge potential
field $A_\mu$ arises as a result of phase fluctuations of $\Delta$; it can then
be argued that $A_\mu$ is most naturally governed by an action of a Maxwell
type \cite{Franz:2002qy,Herbut:2002yq}, 
resulting in massless photon degrees of freedom which have an alternative
interpretation as the Goldstone bosons associated with the condensation of dual
vortices \cite{Kovner:1990pz,Motrunich:2003fz}.

QED$_3$ is a quantum field theory whose study has a long history 
(see \cite{Hands:2002dv} for a brief review).
The main issue is chiral symmetry breaking ($\chi$SB), i.e.
whether chiral symmetry, the invariance of the action
under independent global rotations of left- and right-handed helicity spinors,
is spontaneously broken, signalled by a chiral condensate
$\langle\bar\Psi\Psi\rangle\not=0$. 
$\chi$SB implies dynamical mass generation, i.e.
the physical fermion mass $M$ may be much greater than 
the ``bare'' or Lagrangian mass $m$.
This is believed to depend sensitively 
on the number of fermion species $N_f$ in the model; $\chi$SB is supposed to
occur only for $N_f$ less than some critical $N_{fc}$, whose precise
value remains a goal of non-perturbative quantum field theory.

In the condensed-matter context, the $\chi$SB order parameter can be mapped
directly into the SDW one. If $\chi$SB does not occur (i.e. $N_f>N_{fc}$),
then the resulting theory of light fermion degrees of freedom interacting with 
massless gauge degrees of freedom is proposed as a theory of the so-called
``pseudogap'' region of the cuprate phase diagram, characterised by spectral
depletion in the immediate vicinity of the Fermi energy even in the absence of a
well-defined quasiparticle peak. The main prediction of the QED$_3$ approach
is thus that if $N_{fc}>2$ the dSC and SDW phases are connected in the $T\to0$
limit \cite{Herbut:2002yq}, whereas if $N_{fc}<2$ they are separated 
by a region of pseudogap phase \cite{Franz:2002qy}. 

An important assumption in the above chain of reasoning is that results from 
continuum QED$_3$ in the isotropic limit (as usually studied in quantum
field theory) can be
applied directly to the condensed-matter system, whose Lagrangian density
(\ref{eq:finalbit}) below has kinetic terms describing a single flavour with
differing strengths or 
``velocities'' in $x$- and $y$-directions as an artifact of the transformation
to the relativistic spinor basis. This is in principle not a negligible effect;
the velocity ratio or {\em anisotropy} $\kappa$ in real cuprates varies with $x$
\cite{Sutherland}, and can be as
large as 7 at the onset of the dSC phase \cite{Tallon}.
Evidence in favour of applying predictions of the isotropic theory comes from a
renormalisation-group analysis, which studied small anisotropy perturbations to
the isotropic system and concluded that weak anisotropy is an irrelevant
perturbation \cite{Lee:2002qz,Vafek:2002jf}. This result is then used to argue
that the critical $N_{fc}$ is a universal constant, independent of $\kappa$, and
hence that the various estimates of $N_{fc}$ in the literature can be applied to
the cuprate problem.

It should be  noted here that similar ideas regarding relativistic fermions (often four-Fermi theories) have been also been discussed in the literature relating to graphene and similar compounds, both theoretically (for example, see \cite{Semenoff:1984,Gusynin:cond-mat0506575,Gusynin:cond-mat0411381,Peres:cond-mat0506709}) and experimentally (for example \cite{Novosolev:2005}).  However, this form has only an asymmetry between temporal and spatial directions, whereas in what follows we treat a more generally anisotropic system where all three Euclidean axes are distinguished.  

In our previous paper \cite{Hands:2004ex} we made the first study of anisotropic
QED$_3$ using the methods of numerical lattice gauge theory, a non-perturbative
technique with very different systematic approximations to 
continuum-based approaches. Using a large value of the 
fermion-photon coupling strength, we studied the $\chi$SB order parameter
$\langle\bar\chi\chi\rangle$ on a 
relatively modest $16^3$ spacetime lattice as the anisotropy $\kappa$ is
increased from 1, and provided
preliminary results that suggest the existence of a chiral symmetry restoring
phase transition at a critical value $\kappa_c$; moreover, we found that the 
``renormalised'' $\kappa_R$ -- obtained by considering the spatial decay of
correlations of pseudoscalar meson or ``pion'' fermion -- anti-fermion 
($f\bar f$) bound states
-- obeyed $\kappa_R>\kappa$, suggesting in contrast to
\cite{Lee:2002qz} that $\kappa$ is in fact a {\em relevant} parameter.
Both observations suggest caution should be used in applying isotropic QED$_3$
directly to cuprates.
 
Several questions raised by \cite{Hands:2004ex} are addressed in the current
paper. Firstly, we wish to understand the nature of the chiral symmetry
restoring transition, using the traditional method of studying the transition
on larger systems and applying a finite-volume scaling analysis. New results
on $20^3$ and $24^3$ lattices, together with analyses assuming both
isotropic and weakly anisotropic finite volume scaling are 
presented in Sec.~\ref{sec:fvs}. As we shall see, the data is best fitted by an
ansatz which takes anisotropy into account, which suggests that the value of
$\kappa_c$ in the thermodynamic limit may be considerably larger than the
estimate $\kappa_c\simeq4.5$ of \cite{Hands:2004ex}.
Next, in Sec.~\ref{sec:scalar}
we have studied spectroscopy in another $f\bar f$ channel with
scalar, rather than pseudoscalar, quantum numbers. This is important for two
reasons. Firstly, as the parity partner of the pseudoscalar the scalar should
become degenerate with the pion at large $\kappa$, giving further
evidence for the restoration of chiral symmetry. Secondly, since the pion is the
Goldstone boson associated with $\chi$SB in the low-$\kappa$ phase, it is
in some sense a ``distinguished particle'', and motivates us to check 
the renormalised anisotropy $\kappa_R$ using a different channel.

Finally, for the first time we present results for the fermion propagator
$\langle\chi(x)\bar\chi(y)\rangle$.
Since
this is not a gauge invariant object, in order to obtain a non-zero result 
this has necessitated the implementation 
of a gauge fixing procedure, described in some detail in
Sec.~\ref{sec:gaugefix}.
Our motivation comes from the arguments of Te\v sanovi\'c {\em et al.}
\cite{Franz:2002qy,Franz:2002bc}, suggesting that the 
massless quasiparticles of the pseudogap phase
acquire a small,
gauge dependent anomalous dimension due to their interaction with the
statistical gauge field, which may explain non-standard scaling of 
transport coefficients such as resistivity and thermal conductivity in
the pseudogap phase. From a numerical point of view this has proved easily the
most demanding part of the project, requiring much computational effort to
extract any kind of signal from the statistical noise inherent in the Monte
Carlo method. Somewhat unexpectedly, we find evidence for the persistence of 
a dynamically generated fermion mass in the high-$\kappa$ phase, despite the
apparent restoration of chiral symmetry. A physical scenario consistent with 
these observations is discussed further in Sec.~\ref{sec:conclusion}.

\section{Review of the Lattice Model}

\subsection{QED$_3$ as an effective theory of the pseudogap}
\label{sec:qed3}

The mapping of the pseudogap region of the cuprate phase diagram onto QED$_3$ is
derived in detail in \cite{Franz:2002qy,Herbut:2002yq}, and reviewed in language
more accessible to particle physicists in
\cite{Hands:2004ex}. Here we briefly summarise, starting
with the following Euclidean (imaginary time) action, also known as 
the Bogoliubov -- deGennes model, for $d$-wave quasiparticles in the dSC phase.
\begin{eqnarray}
S&=&T\sum_{\vec k,\sigma,\omega_n}\biggl[(i\omega_n-\xi_{\vec k})
c^\dagger_\sigma(\vec k,\omega_n)c_\sigma(\vec k,\omega_n)\nonumber\\
&-&{\sigma\over2}\left(\Delta(\vec k)c^\dagger_\sigma(\vec
k,\omega_n)c^\dagger_{-\sigma}(-\vec k,-\omega_n)
-\Delta^\dagger(\vec k)c_\sigma(\vec
k,\omega_n)c_{-\sigma}(-\vec k,-\omega_n)\right)\biggr],
\label{eq:BdG}
\end{eqnarray}

\noindent where $c^\dagger,c$ are creation and annihilation operators for electrons with spin $\sigma=\pm1$, $\omega_n=(2n-1)\pi T$ are the allowed Matsubara frequencies, the function $\xi_{\vec k}$ is the energy of a free quasiparticle 
(which thus vanishes for $\vec k$ on the Fermi surface), and $\Delta(\vec k)$ is the gap function, which can be thought of as a self-consistent pairing field.
Due to its $d$-wave symmetry, $\Delta$ actually vanishes at two pairs of node
momenta $\vec k=\pm\vec K_1, \pm\vec K_2$, with $\vec K_1.\vec K_2=0$.

Linearising the latter functions around the nodes and defining the 4-spinor $\Psi_i$ at the node pair $i$ as 
\begin{equation}
\Psi_i^{\mbox{tr}}(\vec q,\omega)=\left(c_+(\vec k,\omega), c^\dagger_-(-\vec
k,-\omega), c_+(\vec k-2\vec K_i,\omega), c^\dagger_-(-\vec k+2\vec
K_i,-\omega)\right),
\end{equation}
we may write the following effective action describing the behaviour
of the system at low $T$ \cite{Lee:2002qz}:

\begin{eqnarray}
S=\int d^2r\int_0^\beta d\tau
\bar\Psi_1[\gamma_0D_\tau&+&\delta\kappa^{-\frac{1}{2}}\gamma_1D_x
+\delta\kappa^{\frac{1}{2}}\gamma_2D_y]\Psi_1+\nonumber\\
\bar\Psi_2[\gamma_0D_\tau&+&\delta\kappa^{-\frac{1}{2}}\gamma_1D_y
+\delta\kappa^{\frac{1}{2}}\gamma_2D_x]\Psi_2 + \frac{1}{2g^2}F_{\mu\nu}^2,\nonumber\\
\left.\right.
\label{eq:finalbit}
\end{eqnarray}
where $\beta\equiv1/T$, 
$\kappa=v_F/v_\Delta$ 
(where $v_F$ and $v_\Delta$ are the {\em Fermi} and {\em Gap}
velocities derived from the linearisation of $\xi_{\vec k}$ and  $\Delta(\vec
k)$ respectively about the nodes) is the anisotropy, $\delta=\sqrt{v_F
v_\Delta}$, and
the $4\times4$ traceless
hermitian matrices $\gamma_\mu$ obey
$\{\gamma_\mu,\gamma_\nu\}=2\delta_{\mu\nu}$.
The action (\ref{eq:finalbit}) describes $N_f=2$ flavours of relativistic
fermion $\Psi$ (sometimes known as `nodal fermions' in this context) 
interacting with an abelian gauge potential $A_\mu$, which we will often refer
to as the `photon', via the covariant derivative
$D_\mu\equiv\partial_\mu+iA_\mu$.
The photon-fermion interaction models the effect of the phase fluctuations of
the pairing field $\Delta$:
photon dynamics are governed by
$F_{\mu\nu}^2\equiv(\partial_{[\mu}A_{\nu]})^2$, and the coupling $g$ (the
analogue of `electron charge' in textbook QED) 
is related to the diamagnetic susceptibility $\chi$ via
$g\sim\chi^{-{1\over2}}$ \cite{Franz:2002qy}.

The two velocities depend on the shape of the Fermi surface, and hence on the
doping of the superconductor \cite{Sutherland,Tallon}, 
implying that the same is true of $\kappa$; at the onset of superconductivity
at low $T$ $\kappa$ may be as much as $O(7)$.

\subsection{Lattice Model of Anisotropic QED$_3$}

The formulation of 
isotropic \qed on a spacetime lattice
is described in detail in \cite{Hands:2004bh}; 
in what follows we summarise the treatment of our anisotropic model 
given in \cite{Hands:2004ex}.
For $N$ flavours of staggered lattice
fermion, the following is a \qed action with an explicit spatial anisotropy:
\begin{equation}
S =\sum_{i=1}^{N} \sum_{x,x'} a^3\bar{\chi_{i}}(x) M_{x,x'} \chi_{i}(x') 
+\frac{\beta}{2} \!\!\sum_{x,\mu<\nu}\!\! a^3\Theta_{\mu\nu}^{2}(x).
\label{eqn:lattact}
\end{equation}
We define the fermion matrix $M_{x,x'}$ as follows:
\begin{equation}
	M_{x,x'} =  {1\over2a}\sum_{\mu=1}^{3} \xi_{\mu}(x) 
[\delta_{x',x+\hat{\mu}} U_{x\mu} -
\delta_{x',x-\hat\mu}U_{x'\mu}^\dagger]
+ m\delta_{\mu\nu}  \label{eqn:fermion_matrix}
\end{equation}
where $\xi_{\mu}$ is
\begin{equation}
\xi_{\mu}(x) = \lambda_{\mu} \eta_{\mu}(x)
\end{equation}

\noindent and $\eta_{\mu}(x)=(-1)^{x_{1} + ... + x_{\mu-1}}$, 
where $x_1=x$, $x_2=y$ and
$x_3=\tau$, is the Kawomoto-Smit phase of the staggered fermion field. 
The physical lattice spacing is denoted by $a$.
The $\lambda_{\mu}$ are anisotropy factors, which we define like so: 
$\lambda_{x}=\kappa^{- \frac{1}{2}}$, $\lambda_{y}=\kappa^{\frac{1}{2}}$, 
$\lambda_{t}=1$.  The $\eta$ factors 
ensure that the action describes relativistic covariant
fermions in the isotropic limit $\kappa=1$.

Taking the photon-like degree of freedom $\theta_\mu(x)$ to exist on the link
connecting site $x$ to site $x+\hat\mu$, makes
$U_{\mu}(x)\equiv \exp(ia\theta_{\mu}(x))$ in (\ref{eqn:fermion_matrix}) 
the parallel transporter defining the gauge interaction with the fermions; 
we may define 
a non-compact gauge action via
\begin{equation}
	\Theta_{\mu\nu}(x) = {1\over a^2}[\Delta_{\mu}^{+}\theta_{\nu}(x)-  
\Delta_{\nu}^{+}\theta_{\mu}(x)]. 
\end{equation}
The dimensionless parameter $\beta$ is given in terms of the QED coupling
constant 
via $\beta\equiv1/g^2a$. It is
convenient to work wherever possible in `lattice units' such that $a=1$.

It is important that a distinction is made between our particular use of
anisotropy, which treats it as a physical property of the system that can be
observed and renormalised through quantum corrections, and the more general use
of anisotropic cutoffs in lattice field theory, wherein the anisotropies are
controlled such that they disappear in the continuum limit, maintaining the
Lorentz covariance of the theory.  In this latter case, anisotropy is not
physically observable.

\subsection{The Simulation}\label{sec:sim}

In Reference~\cite{Hands:2004ex} we simulated the dynamics of the lattice
action (\ref{eqn:lattact},\ref{eqn:fermion_matrix}) using a hybrid Monte Carlo
algorithm on a $16^3$ lattice for
$\kappa$ ranging from 1 to 10 and the bare mass $m=0.05,\ldots,0.01$. 
The gauge coupling constant $\beta$ was held at a
constant value 0.2 throughout -- at this relatively strong coupling the system
is in a state of spontaneously broken chiral symmetry at $\kappa=1$.
The main results of \cite{Hands:2004ex} are that the chiral condensate decreases
with increasing $\kappa$,  consistent with a second order chiral symmetry
restoring transition at $\kappa_c=4.35(2)$, and that the renormalised
anisotropy $\kappa_R$ obtained by comparison of pion correlators in $x$- and
$y$-directions obeys
\begin{equation}
\kappa_R-1\approx2(\kappa-1),
\end{equation}
implying that $\kappa$ is a relevant parameter.

In the calculations presented in this paper,
unless otherwise noted, the gauge configuration ensemble
$\{\theta\}$ used was generated using the same
Hybrid Monte Carlo algorithm, running for around 1000 trajectories of mean 
length 1.0 on $L^3$ lattices with $L=16$, 
and the gauge coupling set to the same value
$\beta=0.2$.
Even-odd partitioning was used; this allowed us to set $N=1$, giving us $N_f=2$ in the continuum limit.  Typical acceptances were $60-70\% $ for $m=0.01$, and $70-80\% $ for other bare mass values.

A novelty of this paper is that we have extended our study to a range of
volumes: 
datasets for the $20^3$ and $24^3$ lattices typically contain 700 and 600-700 trajectories per point with acceptances of $82-94\%$ and $75-82\%$ respectively.  
In the the studies of fermion propagation presented in Sec.~\ref{sec:fermion},
gauge-fixed configurations were generated on a $16^3$ lattice and consisted of 
$\sim$30,000 trajectories per point with an acceptance rate of $79-87\%$.

\section{Susceptibilities and Finite Size Scaling}
\label{sec:fvs}

We begin the presentation of our results with measurements of 
longitudinal susceptibility and the chiral condensate
as $L$ is varied. This is necessary to pin down the
nature of the chiral symmetry restoring transition with more precision. Apart
from the intrinsic theoretical interest, there are important phenomenological
issues at stake. Firstly, it is important to know the value of the critical
anisotropy $\kappa_c$ at which the transition takes place in the continuum and
thermodynamic limits, since in principle this is a physically observable
parameter in real cuprates~\cite{Sutherland}. Secondly, the order of the phase
transition is important; were it either first-order or a crossover, 
then an immeasurably small but
non-vanishing condensate may persist in the high-$\kappa$ ``chirally restored
phase'', meaning that antiferromagnetic order can survive the
transition~\cite{Hands:2004ex}. As we shall see below, the results we have been
able to obtain with our resources have not settled the issue unequivocally;
it seems likely that a model of finite volume scaling which takes account of 
the anisotropy is
required.

\subsection{Finite size scaling of the condensate}

Here we present the results of a preliminary study of the finite size scaling 
of the chiral condensate and longitudinal susceptibility at fermion
mass $m=0.01$, the smallest of the bare masses examined in \cite{Hands:2004ex}.
We define the chiral condensate in terms of the trace of the inverse
of the fermion matrix $M$:
\begin{equation}
\langle\bar{\chi}\chi\rangle=-{1\over V}{{\partial\ln Z}\over\partial m}=
{1\over V}\langle\mbox{tr}M^{-1}\rangle,
\end{equation}
and the longitudinal susceptibility in terms of its derivative,
\begin{equation}
\chi_{l}=\frac{\partial\langle\bar{\chi}\chi\rangle}{\partial m}
={1\over
V}[\langle(\mbox{tr}M^{-1})(\mbox{tr}M^{-1})\rangle-
\langle\mbox{tr}M^{-1}\rangle^2-\langle\mbox{tr}(M^{-1}M^{-1})\rangle].
\label{eq:lsusc}
\end{equation}
Note that eqn. (\ref{eq:lsusc}) includes diagrams which
are both connected and disconnected in terms of fermion lines; 
both contributions were calculated.

In the vicinity of the phase transition
$\chi_l$ should peak at an anisotropy $\kappa_{peak}$ which should
tend towards the critical value $\kappa_c$ in the thermodynamic limit.
Examining the plot of the longitudinal susceptibility as the size of the lattice
is varied (Figure \ref{fig:scallongsus}), we observe that the peak shifts to the
right by an amount that decreases as the lattice size increases; this suggests
that a second order transition might occur at a  finite value of $\kappa_c$  in
the thermodynamic limit.  Unexpectedly, however, the magnitude of the peak
appears
suppressed as the lattice size increases.  This may have several possible causes:

\begin{figure}[htb]
\vspace{.5cm}
\begin{center}
\epsfig{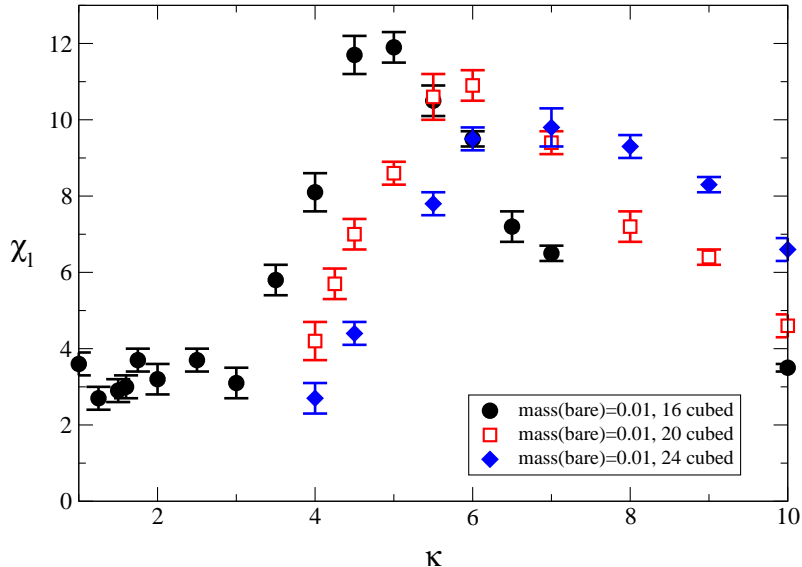}
\end{center}
\vspace{-.5cm}
\caption{\small $\chi_l$ for various lattice sizes and $m=0.01$.}
\label{fig:scallongsus}
\end{figure}

\begin{itemize} 
\item The magnitude of the peak does increase, but the width of the peak as the
lattice volume increases narrows such that it falls between the available  data
points and is not detected.  The rounded shape of the curves suggests that 
this is unlikely.  
 
\item  This is not a second order phase transition; perhaps we're observing a crossover instead. 
If there is a crossover between the two phases, and not a genuine
second-order phase transition, then a small chiral condensate is expected to 
persist in the high-$\kappa$ phase.  Some analytic approaches 
predict dimensionless condensates
$\beta^2\langle\bar\chi\chi\rangle$ as small as
$O(10^{-4})$~\cite{Appelquist:2004ib}.  
Attempts to rule out this possibility
regarding the chiral phase transition in studies of isotropic \qed with
various $N_f$ such as \cite{Hands:2004bh,Hands:2002dv} have not proven to be 
successful, and it is also likely to be as difficult in this case.
 
\item Our system has an anisotropic coupling between the gauge and fermion
fields.  The effects of this could be difficult to account for in the standard
finite size scaling developed for
phase transitions in isotropic systems; we should turn our
attention to the scaling of anisotropic systems instead. In the statistical
mechanics literature, one observes two models of this scaling: 
	\begin{itemize}
	\item[]{\bf Weak anisotropy:}  In these systems, there exist different
correlation lengths in different directions; these correlation lengths can be
rescaled such that the system is effectively isotropic in the scaling region
(\cite{chen-2004-} and references therein, notably
\cite{Privman1984,Indekeu1986,Hucht2002}).  We examine this possibility in
detail below.

	\item[] {\bf Strong anisotropy:} In these systems in addition to
correlation lengths, the critical exponent $\nu$ is different in different
directions (\cite{Hucht2002,Caracciolo2003} and references therein).  The
scaling behaviour of these systems  is very sensitive to the shape of the
lattice, and is difficult to treat with data generated on cubic lattices; 
it is mentioned here as an issue worthy of further investigation. 
	\end{itemize}
\end{itemize}

Since our data is restricted to that generated on a square lattice, we will examine only weak anisotropic scaling as compared to isotropic scaling.

\subsubsection{Isotropic scaling}

Firstly, we shall discuss the scaling of the system if it is assumed intrinsically isotropic.
As argued in  \cite{DelDebbio:1997dv}, we can use the scaling behaviour of the system as we vary $L$ in order to determine how the finite volume affects the equation of state.  We do this by treating the inverse linear size of the lattice, $L^{-1}$, as an irrelevant scaling field and use the following as our ansatz, where $k=(\kappa-\kappa_c)$:
\begin{equation}
m=B\chibchi^\delta+ A(k+CL^{-1/\nu})\chibchi^{\rho}. \label{eqn:fscl}
\end{equation}
Here $\delta$, $\nu$ and $\beta\equiv(\delta-\rho)^{-1}$ have their usual
meanings as critical indices describing a continuous phase transition.

\subsubsection{Weakly anisotropic scaling}

In this case, we wish to account for the distortion of the correlation
lengths of the system along the $x$ and $y$ axes by  $\kappa\neq1$.  Finite size
effects enter into the scaling whenever 
\begin{equation}
\xi_\mu\gg L_\mu, \label{eqn:corrscal}
\end{equation}
where $\xi_\mu$ is the correlation length in the direction $\mu$ and $L_\mu$ is
the length of the lattice in that direction.  We introduce three irrelevant
scaling fields: $L^{-1}_1$,  $L^{-1}_2$ and $L^{-1}_3$, defining them in terms
of $L$, the number of lattice spacings along one dimension of the system,
by rescaling $\xi_\mu$ \cite{chen-2004-,Privman1984,Indekeu1986,Hucht2002} such that 
\begin{equation}
\xi^{re}_1=\xi^{re}_2=\xi^{re}_3
\end{equation}
Since (to a first approximation) $\xi^{re}_1=\sqrt{\kappa}^{-1}\xi$, 
$\xi^{re}_2=\sqrt{\kappa}\xi$ and $\xi^{re}_3=\xi$ 
(where $\xi$ is the correlation length of the isotropic system), this gives
\begin{equation}
\sqrt{\kappa}{\xi^{re}_{1}}=\sqrt{\kappa}^{-1}\xi^{re}_{2}=\xi^{re}_{3}.
\end{equation}
This rescaling of the correlation lengths is equivalent to resizing the lattice thus (from consideration of (\ref{eqn:corrscal})):
\begin{equation}
L_1=\sqrt{\kappa}L, \; L_2=\sqrt{\kappa}^{-1}L, \; L_3 =L. 
\label{eq:els}
\end{equation}
So, we can write:
\begin{eqnarray}
\mathcal{V}_{effects}&=&C\left(\frac{1}{L_3}\right)^{1/\nu}+D\left(\frac{1}{L_2}\right)^{1/\nu}+E\left(\frac{1}{L_1}\right)^{1/\nu}
\nonumber\\
&=& C\left(\frac{1}{L}\right)^{1/\nu}+D\left(\frac{\sqrt{\kappa}}{L}\right)^{1/\nu}+E\left(\frac{1}{\sqrt{\kappa}L}\right)^{1/\nu}\nonumber\\
&\equiv&R(\kappa;C,D,E)L^{1/\nu}.
\end{eqnarray}
This motivates the replacement
\begin{equation}
\frac{C}{L^{1/\nu}}\to R(\kappa;C,D,E)L^{-1/\nu}
\end{equation}
in (\ref{eqn:fscl}), which we may then use to study the scaling if weak anisotropy is assumed.

\subsubsection{Results and Discussion}

 We should note that the above equations are only good descriptions of the
behaviour of the system near to a continuous phase transition.  
We have attempted fits to the finite-volume equation of state (\ref{eqn:fscl})
using data from $16^3$, $20^3$ and $24^3$ with $m=0.01$.
To ensure stability of the fit we found that it was also necessary
to include the $m=0.02$ data for the $16^3$ lattice, presented in \cite{Hands:2004ex}, giving 34 data points in all. 

In addition, in order to increase the tractability of our fits, we have made use of the following hyperscaling relation (with dimensionality set to 3):
\begin{equation}
\nu=\frac{(\delta+1)}{3(\delta -\rho)} \label{eqn:nuhypscl}
\end{equation}
which reduces the number of free parameters in our fit to six assuming
isotropic scaling and eight assuming weakly anisotropic.

Results from fitting the chiral condensate data to (\ref{eqn:fscl})
are 
shown in Table \ref{tab:fscaleos}.  
The daggered quantities were obtained through the following relations:
\begin{equation}
\delta = \frac{5-\eta}{1+\eta};\;\;\;
\beta=\frac{1}{2}\nu(1+\eta);\;\;\;
\rho=\delta- \frac{1}{\beta}.
\end{equation}

\begin{table}[htb]
\centering
\setlength{\tabcolsep}{1pc}
\renewcommand{\arraystretch}{1.1}
\protect\footnotesize{
\begin{tabular}{|c|c|c|c|}
\hline
Quantity & Isotropic &  Weakly anisotropic \\
\hline
$A$ &.0393(8)  & .0111(7)  \\
$B$ & 1.28(8) & 1.02(6) \\
$C$ & 368(16)  & -755(338)\\
$D$ & -- 	& 527(78)\\
$E$ & -- 	& -716(447)\\
$\kappa_c$ & 7.66(5)  & 12.3(6)\\
$\delta$ & 3.40(6) & 3.33(6)\\
$\rho$ & .991(7) & 1.01(1)\\
$\beta^\dag$ & .41(1) & .433(3)\\  
$\eta^\dag$ & .363(6) &.386(7)\\
$\nu^\dag$ & .61(2)   &.62(2)\\
$\frac{\chi^2}{d.o.f.}$ & 162  & 6 \\
\hline
\end{tabular}}
\caption{\small Equation of state fit results, allowing for finite size scaling.  Daggered values are calculated from hyperscaling relations (see main text).}
\smallskip
\label{tab:fscaleos}
\end{table}

\begin{figure}[htb]
\vspace{.5cm}
\begin{center}
\epsfig{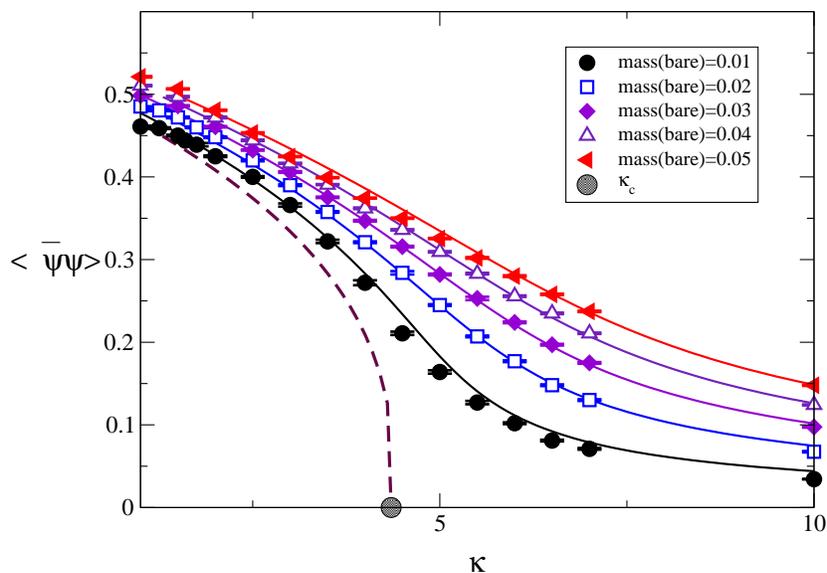}
\end{center}
\vspace{-.5cm}
\caption{\small The corrected plot of the chiral condensate and the equation of
state fits on a fixed volume $16^3$.}
\label{fig:correction}
\end{figure}

\begin{figure}[htb]
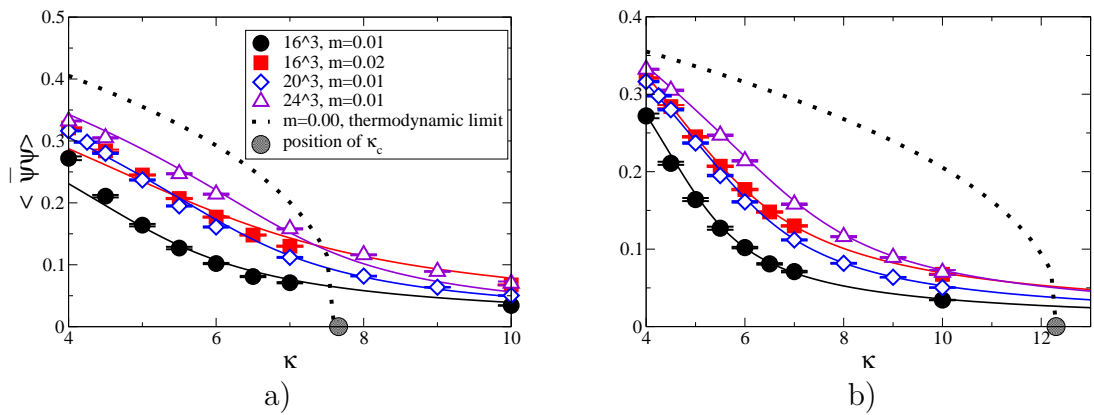

\vspace{1cm}
\begin{center}
\epsfig{file=graphs/isoscl.eps, height=4.8cm}
\hspace{1cm}
\epsfig{file=graphs/anisofit.eps,height=4.8cm}
\end{center}
\vspace{-.5cm}
\hspace{3.8cm}a)\hspace{7.4cm}b)
\vspace{-.2cm}
\caption{\small Equation of state fits for (a) the isotropic case, and (b) the anisotropic case on various lattice sizes and in the thermodynamic, zero mass limit. $16^3$ results are taken from \cite{Hands:2004ex}.}
\label{fig:scaleos}
\end{figure}

The equation of state fits found on 16$^3$
are plotted in Fig.~\ref{fig:correction} 
\footnote{in fact, the original figure shown in
\cite{Hands:2004ex} had incorrect curves, and $\kappa_c$
was not located correctly, although the values of the critical exponents given 
were correct, and the conclusions of that paper remain unaffected.};
for comparison 
the new equations of state, together with the fitted data and the
extrapolation to the chiral limit $m\to0$, are plotted in
Fig.~\ref{fig:scaleos}.
The following features are perhaps the most intriguing.  The
critical indices are compatible for both fits -- however, 
the value of $\kappa_c$ is not only different from the value
$\kappa_c=4.35(2)$ derived from fitting to the $16^3$ data
alone~\cite{Hands:2004ex}, but is significantly different between the two forms of the finite scaling fit.  This suggests not only that finite size effects play a significant role in the behaviour of this system, but that the effects of the anisotropy should be taken into consideration in future studies of the system.  It is also worth noting that the value of the ${\chi^2}/{d.o.f.}$ is significantly better for the anisotropic scaling.
Note also that for the weakly anisotropic fit, the sign of the coefficient of
$L_2$, $D$ is different from those of $L_1$ and $L_3$, $C$ and $E$.  
This may reflect the expectation following (\ref{eq:els})
that $\xi_2\gg L$ over a much wider region of
$\kappa$ than is the case for the other two directions.  {\em Contra} the
fitting results of \cite{Hands:2004ex}, which were confined to a single lattice
size, $\rho\approx 1.00$ with both equations; however, it is plausible that this is due to an insufficient spread of mass
values in the data set.

Whether the actual value of the weakly anisotropic $\kappa_c$ is in fact 12.3(6)
seems  doubtful; we must note that the extrapolation is well outside the region
of $\kappa$ for which we have any data.  An interesting possibility is that it
could also indicate that there is no phase transition and that the fit could be
attempting to compensate for its absence by giving it a value in the unexplored
region. If this behaviour were to persist for a more
extensive data set, this hypothesis could be validated.

\section{Scalar Sector} 
\label{sec:scalar}

The scalar meson is the parity partner of the pseudoscalar pion bound state
studied in \cite{Hands:2004ex}. In a phase with broken chiral symmetry the pion
is a Goldstone boson, and hence is much lighter than the scalar. One signal for
restoration of chiral symmetry is the recovery of degeneracy between scalar and
pseudoscalar in the $m\to0$ limit.
The propagator of the scalar is defined in terms of the fermion fields
as follows:
\begin{equation}
C_{\sigma\mu}(x_\mu)=\sum_{\nu\not=\mu}\sum_{x_\nu}
\langle\bar\chi\chi(0)\bar\chi\chi(x)\rangle.
\label{eq:scalcorr}
\end{equation}

Due to the nature of the flavour structure of staggered lattice fermions,
propagation in this channel
is prone to mixing with low mass bound states with different spin quantum
numbers
\cite{DelDebbio:1997dv}.  Where this contamination is significant, the 
propagator takes on a sawtooth shape, and we must thus fit a four parameter
function, such as that in (\ref{eqn:sclsaw}) below, so that we can 
distinguish propagation in the channel of interest. 

In the following we distinguish between propagation in the Euclidean time
direction $\tau$, 
yielding information on the excitation spectrum in the channel in
question, eg. the bound-state mass, and propagation in the spatial directions
$x$, $y$, where the corresponding quantity is the inverse screening length.
Of course, in an isotropic system the two cases are equivalent in the infinite
volume limit.

\subsection{Temporal propagator}

\begin{figure}[htb]
\vspace{0.5cm}
\begin{center}
\epsfig{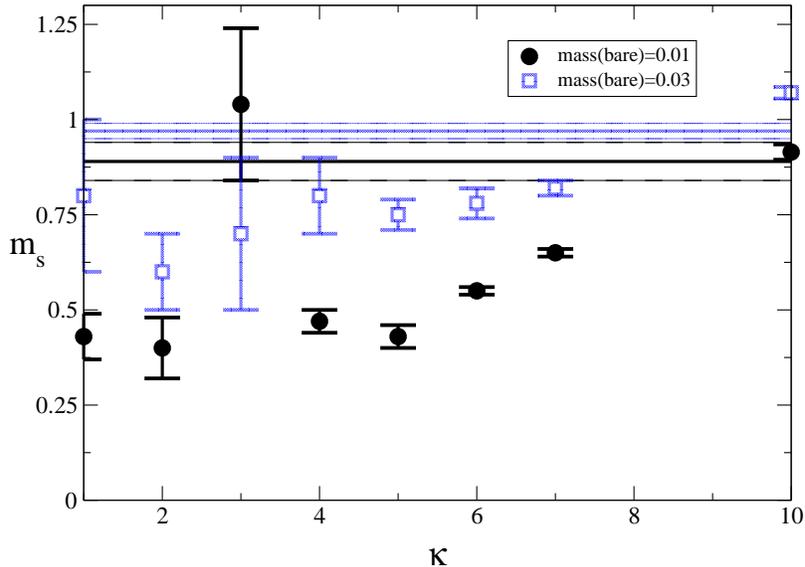}
\end{center}
\vspace{-.5cm}
\caption{\small Scalar masses in the $\tau$ direction. Straight lines represent the pion masses at $\kappa=10.00$, taken from \cite{Hands:2004ex}.  $m=0.05$ values are omitted due to the size of their error bars.}
\label{fig:tscal}
\end{figure}

Least squares fitting of the function 
\begin{equation}
C_{\sigma\,\mu}(x_\mu)=A(e^{- m_{\sigma\,\mu}x_\mu} + e^{-m_{\sigma\,\mu} 
(L_{\mu} -x_\mu)}),
\label{eqn:cosh}
\end{equation}
(with $\mu$ chosen to be $\tau$)
to data from $16^3$ lattices 
proved to be difficult within the chirally broken phase
-- the propagator data was exceedingly noisy, and care had to be taken in order
to isolate the ground state signal from the excited states --  
but as the values moved into the chirally restored region the procedure became
easier to perform.  The results are listed in Table \ref{tab:scal}, and plotted
in the graph \ref{fig:tscal}, alongside the pion masses of 
\cite{Hands:2004ex} for each bare mass at $\kappa=10.00$.  

\begin{figure}[htb]
\vspace{0.5cm}
\begin{center}
\epsfig{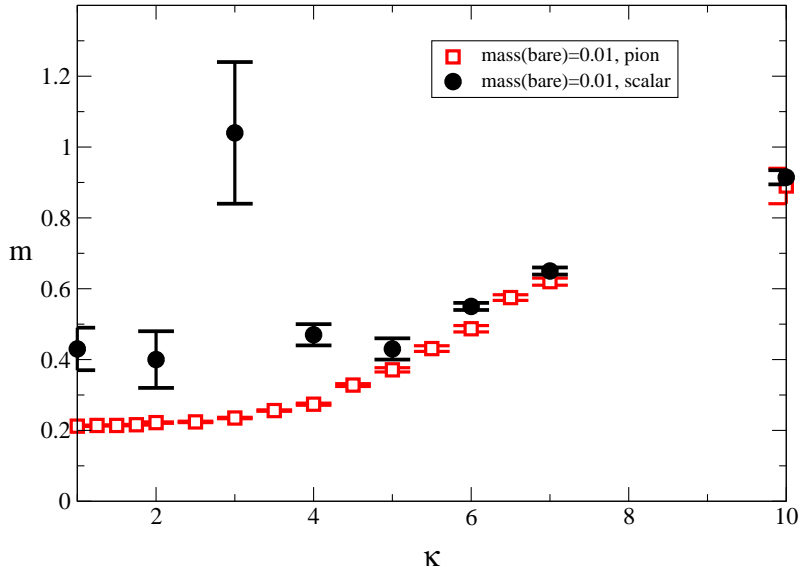}
\end{center}
\vspace{-.5cm}
\caption{\small Scalar and pion masses (from \cite{Hands:2004ex}) in the $\tau$ direction for $m=0.01$. }
\label{fig:comtscal}
\end{figure}

It can be seen from the figure that there are two regimes of scalar behaviour;
below $\kappa_c$, where fitting is quite difficult,
$m_\sigma$ is more or less
constant as $\kappa$ increases (if we go by the $m=0.01$ data and ignore 
the outlier at $\kappa=3.00$) up to
$\kappa\approx5$ (ie. $\kappa\approx\kappa_c$ as estimated on $16^3$), 
whereupon we find that $m_\sigma$ begins to converge with $m_\pi$ as $\kappa$
increases into the chirally restored region.  Figure \ref{fig:comtscal} shows
this in more detail for $m=0.01$. 

 We should point out that the jump in the value of $m_{\sigma\,\tau}$ at
$\kappa=3.00$ and $m=0.01$ (and likely that of $m_{\sigma\,y}$ at $\kappa=4.00$,
$m=0.05$, see below) is likely to be due to the frequent occurrence of abnormally small eigenvalues of the Dirac operator that overlapped with the meson source during our measurement of the propagator, similar to that seen in the Thirring model simulations of \cite{DelDebbio:1999xg}.
 The overall trend is consistent with the scalar becoming degenerate with the
pion at large $\kappa$, consistent with manifest chiral symmetry. 
There is thus no evidence
for persistence of chiral symmetry breaking at large $\kappa$
from the light meson spectrum.

\begin{table}[htb]
\centering
\setlength{\tabcolsep}{1pc}
\renewcommand{\arraystretch}{1.1}
\protect\footnotesize{
\begin{tabular}{|c|c|c|c|c|c|}
\hline
$m$	&$\kappa$ 	&$m_{s}$	&$\chi^2/d.o.f.$	& fit window\\
\hline
0.01&	1.00&	0.43(6)&	1.991&	2-14	\\
&	2.00&	0.40(8)&	1.882&	2-14	\\
&	3.00&	1.04(2)&	1.323&	1-15	\\
&	4.00&	0.47(3)&	1.045&	1-15	\\
&	5.00&	0.43(3)&	1.144&	5-11	\\
&	6.00&	0.55(1)&	1.119&	1-15	\\
&	7.00&	0.65(1)&	0.339&	1-15	\\
&	10.00&	0.91(2)&	0.984&	3-13	\\
\hline
0.03&	1.00	&0.8(2)	&	1.177&	2-14	\\
&	2.00	&0.6(1)	&	1.089&	2-14	\\
&	3.00	&0.7(2)	&	0.923&	2-14	\\
&	4.00	&0.8(1)	&	0.397&	2-14	\\
&	5.00	&0.75(4)&	1.351&	2-14	\\
&	6.00	&0.78(4)&	0.983&	3-13	\\
&	7.00	&0.82(2)&	0.505&	1-15	\\
&	10.00	&1.07(2)&	1.273&	1-15	\\
\hline
0.05&	1.00&	1.0(2)&	2.519	&2-14	\\
&	2.00&	1.2(7)&	1.036	&2-14	\\
&	3.00&	0.6(1)&	1.806	&2-14	\\
&	4.00&	1.0(2)&	0.745	&2-14	\\
&	5.00&	1.4(1)&	0.739	&2-14	\\
&	6.00&	0.95(5)&	1.376	&2-14	\\
&	7.00&	1.05(5)&	0.635	&2-14	\\
&	10.00&	1.16(2)&	0.662	&1-15	\\
\hline
\end{tabular}}
\caption{\small Scalar masses $m_{sc\,\tau}$ in the $\tau$ direction on a $16^3$ lattice, for various masses. }
\smallskip
\label{tab:scal}
\end{table}

\subsection{Spatial propagators}

\begin{table}[htb]
\centering
\setlength{\tabcolsep}{1pc}
\renewcommand{\arraystretch}{1.1}
\protect\footnotesize{
\begin{tabular}{|l|l|l|l|c|}
\hline
$m$ & $\kappa$ & $m_{s x}$ & $\chi^2/d.o.f$ & fit window  \\
\hline
\hline
0.01&1.00	&0.32(4)	&1.052	&2-14	\\
&2.00		&1.0(2)		&0.856	&1-15	\\
&3.00		&2(1)		&0.874	&1-15	\\
&4.00		&1.04(8)	&0.645	&1-15	\\
&5.00		&1.28(3)	&1.215	&1-15	\\
&6.00		&1.53(2)	&1.153	&1-15	\\
&7.00		&1.83(2)	&0.734	&1-15	\\
&10.00		&2.62(2)	&0.55	&1-15	\\
\hline
0.03&1.00	&0.7(2)&	1.643	&2-14	\\
&2.00		&1.2(2)&	1.495	&1-15	\\
&3.00		&1.7(7)&	1.021	&1-15	\\
&4.00		&1.8(3)&	0.577&	1-15	\\
&5.00		&2.3(2)&	1.09	&1-15	\\
&6.00		&2.2(1)&	1.596	&1-15	\\
&7.00		&2.3(1)&	0.984	&2-14	\\
&10.00		&2.80(3)&	0.853	&1-15	\\
\hline
0.05 &1.00		&0.7(1)&	2.485	&2-14	\\
&2.00		&1(1)	&0.47	&2-14	\\
&3.00		&2.2(4)&	0.981	&1-15	\\
&4.00		&3(3)	&0.967	&1-15	\\
&5.00		&2.5(4)&	0.842	&1-15	\\
&6.00		&2.6(2)&	0.637	&1-15	\\
&7.00		&2.8(2)&	0.906	&1-15	\\
&10.00		&3.20(7)&	1.029	&1-15	\\
\hline
\end{tabular}}
\caption{Effective scalar mass $m_{s\; x}$ in the $x$ direction. }
\smallskip
\label{tab:xscalmass}
\end{table}

The spatial scalar masses were also obtained by least squares fitting to the
propagator.  For the $x$-direction (Table \ref{tab:xscalmass}) and the
$\kappa=1.00$ $y$-direction correlation functions, we used the fit function
(\ref{eqn:cosh}), and selected the fit window so as to exclude higher 
mass states.
For $\kappa>1.00$, the $y$-direction correlation function exhibits a saw-tooth
behaviour, motivating the following fit:
\begin{equation}
C_{\sigma\,y}=A(e^{-m_{\sigma\,y}y}+e^{-m_{\sigma\,y}(L_y-y)})
+(-1)^{y}B(e^{-My}+e^{-M(L_y-y)}), \label{eqn:sclsaw}
\end{equation}
which proved acceptable across the full range of $\kappa$ if a fixed fitting
window of spaceslices 1-15 was used.
As with the temporal scalar propagators, those in the chirally symmetric phase
were easier to fit than those in the chirally broken phase.  It is also worth
noting that in the symmetric phase the value of the
correction mass $M$ was often consistent with zero for $m=0.01$ and $m=0.03$.

\begin{table}[htb]
\centering
\setlength{\tabcolsep}{1pc}
\renewcommand{\arraystretch}{1.1}
\protect\footnotesize{
\begin{tabular}{|l|l|l|l|c|}
\hline
$m$ & $\kappa$ & $m_{\sigma y}$ & $\chi^2/d.o.f$ & fit window  \\
\hline
\hline
0.01&1.00	&0.41(6)		&0.56	&2-14\\
&2.00	&0.25(5)		&1.469  & 1-15\\
&3.00	&0.2(1) 		&1.923	&1-15\\
&4.00	&0.13(2) 	&1.548	&1-15\\
&5.00	&0.09(2)		&2.142	&1-15\\
&6.00	&0.1(2)		&1.227	&1-15\\
&7.00	&0.1(5)		&0.608	&1-15\\
&10.00	&00(11) 		&1.735	&1-15\\
\hline
0.03&1.00	&0.8(2)		&0.721	&2-14\\
&2.00	&0.59(7)		&1.732&	1-15\\
&3.00	&0.7(3)		&1.076	&1-15\\
&4.00	&0.50(6)		&7.661	&1-15\\
&5.00	&0.15(4)		&1.015	&1-15\\
&6.00	&0.12(9)		&1.351	&1-15\\
&7.00	&0.10(8)		&0.798	&1-15\\
&10.00	&0.1(2) 		&1.55	&1-15\\
\hline
.05&1.00	&0.9(2)		&0.676	&2-14\\
&2.00	&2.9(6)		&8.468	&1-15\\
&3.00	&0.8(3)		&4.308	&1-15\\
&4.00	&0.28(7)		&1.315	&1-15\\
&5.00	&0.19(5)		&2.097	&1-15\\
&6.00	&0.15(4)		&1.636	&1-15\\
&7.00	&0.12(5)		&1.818	&1-15\\
&10.00	&0.1(3)		&1.87	&1-15\\
\hline
\end{tabular}}
\caption{Effective scalar mass $m_{\sigma y}$ in the $y$ direction. }
\smallskip
\label{tab:yscalmass}
\end{table}

\begin{figure}[htb]
\vspace{0.5cm}
\begin{center}
\epsfig{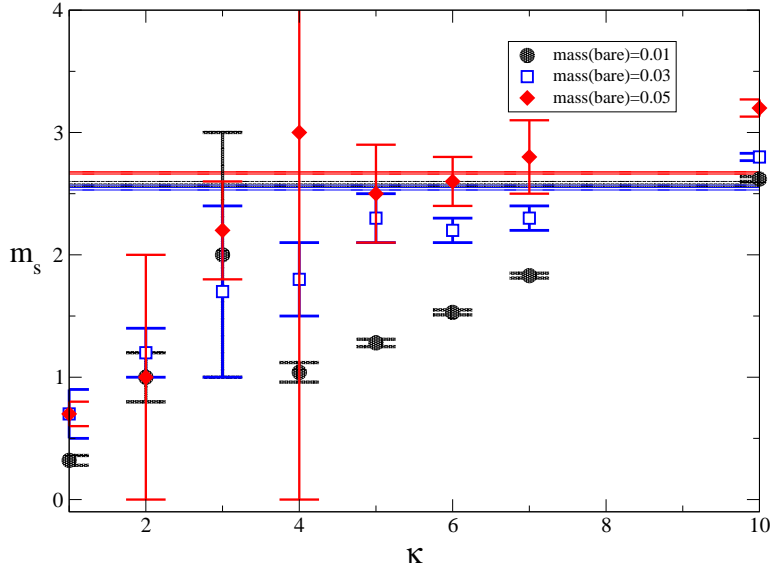}
\end{center}
\vspace{-.5cm}
\caption{\small The scalar screening mass in the $x$-direction, $m_{s\;x}$, on a $16^3$ lattice.  The straight lines represent $m_{\pi\,x}$ at a value of $\kappa=10.00$.}
\label{fig:xscal}
\end{figure}

\begin{figure}[htb]
\vspace{0.5cm}
\begin{center}
\epsfig{file=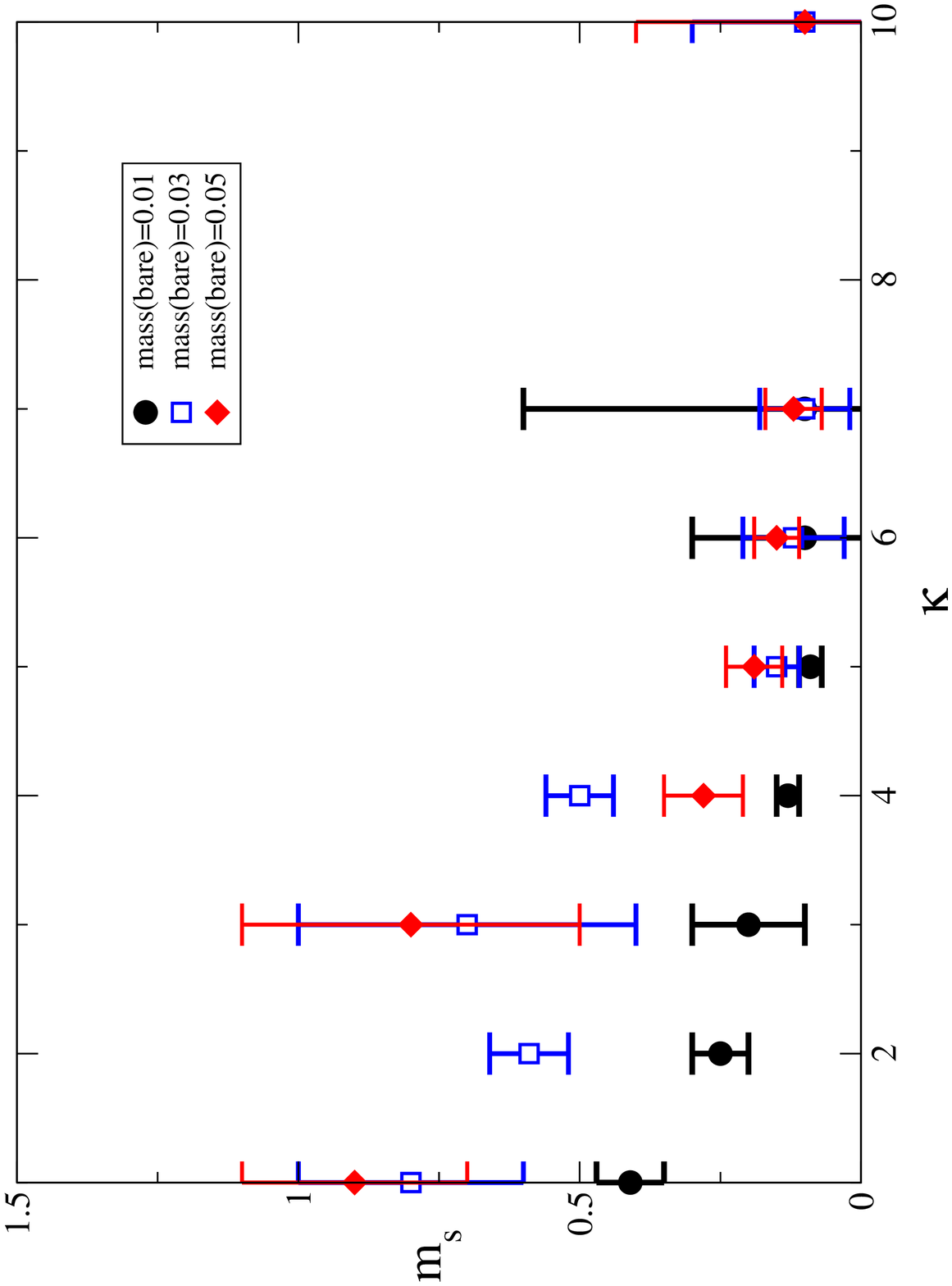, height=7.5cm}
\end{center}
\vspace{-.5cm}
\caption{\small The scalar screening mass in the $y$-direction, $m_{s\;y}$, on a $16^3$ lattice.  We omit the anomalous $m=0.05$, $\kappa=2.00$ value since its inclusion obscures the general trend of the data, and the $m=0.01$, $\kappa=10.00$ value due to the size of its error bars.}
\label{fig:yscal}
\end{figure}

We have plotted $m_{\sigma\;x}$ against $\kappa$ in Figure \ref{fig:xscal}, and
$m_{\sigma\;y}$ against $\kappa$ in Figure \ref{fig:yscal}.  The trends
previously observed for pions in \cite{Hands:2004ex} are repeated here:   the
value of $m_{\sigma\,x}$ increases with $\kappa$. In addition,  as we have seen
with $m_{\sigma\,\tau}$, it appears that there is convergence between the
$m_{\sigma\,x}$ and the $m_{\pi\,x}$ values, most notably for $m=0.01$ (compare
$m_{\sigma\,x}=.32(4)$ and $m_{\pi\,x}=.211(1)$ at $\kappa=1.00$ with
$m_{\sigma\,x}=2.62(2)$ and $m_{\pi\,x}=2.58(2)$ at $\kappa=10.00$).  Just as in
Fig.~\ref{fig:tscal}, there is a  change in behaviour around
$\kappa\approx5$, suggesting a change in behaviour as the scalar masses begin to
converge on the pion masses within the chirally symmetric phase, once again
consistent with the pion and scalar being parity partners.

\begin{figure}[htb]
\vspace{0.5cm}
\begin{center}
\epsfig{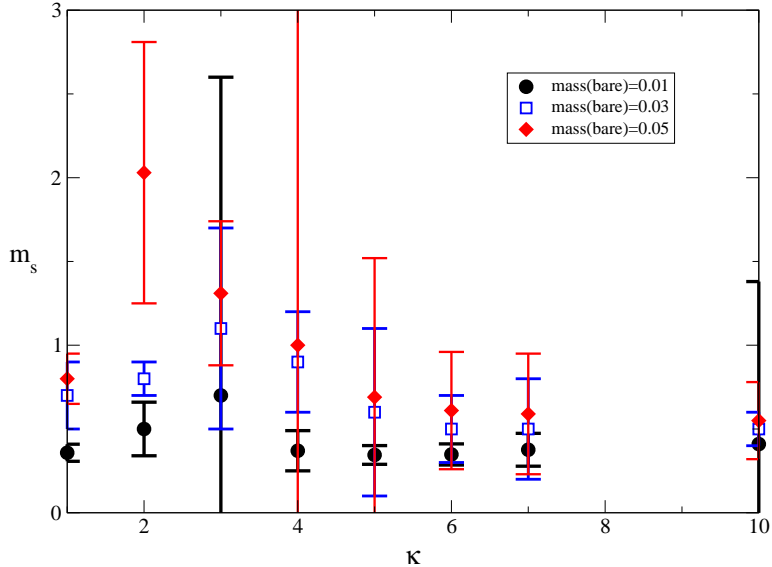}
\end{center}
\vspace{-.5cm}
\caption{\small The geometric mean of the scalar screening masses, $\sqrt{m_{s\;y}m_{s\;x}}$, on a $16^3$ lattice.}
\label{fig:geoscal}
\end{figure}

In the case of the $y$-direction masses, there is no pion data within the
chirally symmetric phase with which to compare our results.  The quality of
the scalar data is also not that good, for the reasons mentioned above.  
It is less clear whether the change in the behaviour 
between phases is present here; it is likely to be quite small, and in any case
the errors easily obscure it.  Our only conclusion is, then, that the value of
$m_{\sigma\;y}$ decreases as we increase $\kappa$; this is reinforced by the
behaviour of the geometric mean of $m_{\sigma\;y}$ and $m_{\sigma\;x}$ above 
$\kappa\approx5\approx\kappa_c$,
(Fig. \ref{fig:geoscal}); while there is a slight increase as we approach
$\kappa_c$ (though the large error bars make it difficult to determine to what
extent this is a genuine effect), above it, the geometric mean appears to remain fairly constant, which implies that $m_{\sigma\;y}$ is decreasing, as $m_{\sigma\;x}$ is increasing in this region.  

\subsection{Renormalised Anisotropy}
\label{sec:krenscal}

\begin{figure}[htb]
\vspace{0.5cm}
\begin{center}
\epsfig{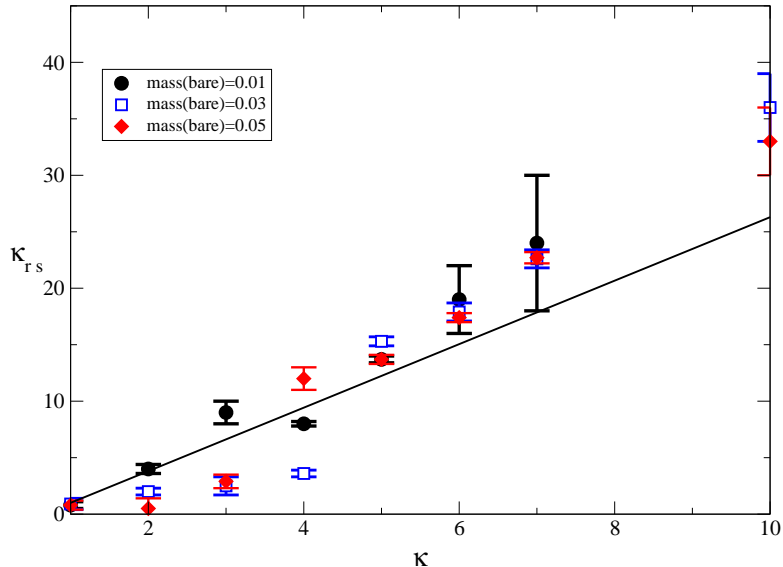}
\end{center}
\vspace{-.5cm}
\caption{\small The renormalised $\kappa$ with respect to scalars,
$\kappa_{R\,\sigma}$, on a $16^3$ lattice, together with a linear fit for $m=0.01$ (the dashed line is the quadratic fit, the filled line the linear). $\kappa=10.00$ has been omitted due to the size of the error bars.}
\label{fig:scalkrenormed}
\end{figure}

Taking $\kappa_{R\,\sigma}$ to be the ratio of $m_{\sigma\,x}$ and
$m_{\sigma\,y}$ (Figure \ref{fig:scalkrenormed}), we find that it is relevant
above $\kappa_c$ (more so, in fact, than for pions; Cf. Fig. 8 of
\cite{Hands:2004ex}); however, in contrast to the pion case 
there appears to be a clear mass dependence below $\kappa_c$. 
It is difficult to tell
whether this is a real effect or merely an artefact of the propagator fitting.
On the assumption that the behaviour for $m=0.01$ is more or less linear, we fitted the data for $1.00\leq\kappa\leq7.00$ to
\begin{equation}
R_\sigma=\frac{(\kappa_{R\,\sigma}-1)}{(\kappa-1)}, 
\end{equation}
and acquired $R_\sigma=2.8(1)$, with $\chi^2/d.o.f.=1.77$. This appears slightly
larger than $R_\pi\simeq2.1$ \cite{Hands:2004ex}, 
suggesting that the behaviour of scalar particles is affected by the anisotropy to a greater extent than that of the pions.


More data is needed before we can make definitive statements.  In addition, it
should be  noted that we can't rule out the existence of a change in behaviour
around $\kappa_c$ for the values of $\kappa_{R\,\pi}$ -- the pion data 
of \cite{Hands:2004ex} doesn't extend far enough. 
Based on the parity partnership of pions and scalars, we propose the following
hypothesis: that in the chirally restored phase, the magnitude of
$\kappa_{R\,\sigma}$ will gradually approach that of $\kappa_{R\;\pi}$.  The
required data would best be generated on considerably larger lattices, with 
better statistics and perhaps with improved operators in order to avoid some of the issues with the data examined here.

\section{Fermion Sector}
\label{sec:fermion}

In this section we report for the first time on studies of the fermion
propagator $\langle\chi(x)\bar\chi(y)\rangle$. Large non-perturbative 
corrections to
this Green function in the chiral limit $m\to0$ have been proposed as an 
explanation of non-Fermi liquid behaviour in the non-superconducting region of
the cuprate phase diagram \cite{Franz:2002qy}. An important challenge,
both technical and conceptual, which must be faced is that the fermion
propagator in QED 
is not a gauge invariant object, and can only by calculated, either
analytically or numerically, if a gauge-fixing procedure is specified
\cite{Gockeler:1991bu}. 
The dependence of the results on the choice of gauge is a thorny issue
\cite{Khveshchenko:2001jn,Khveshchenko:2002xc,Franz:2002bc,Khveshchenko:2002ra};
here we will content ourselves with specifying Landau gauge, ie.
$\partial_\mu A_\mu=0$ in continuum notation (implying that only transverse
degrees of freedom are retained in the photon propagator), and performing
a fully non-perturbative calculation on a $16^3$ lattice. In what follows we
will first devote some considerable attention on the technicalities of fixing 
an unambiguous gauge for lattice gauge fields $U_\mu$, and then report our
results for the fermion propagator. Our strategy in this exploratory study is to
calculate the physical (ie. renormalised) fermion mass $m_f$ for fixed bare mass
$m$ as a function of the anisotropy parameter $\kappa$. Apart from the fact that
this is the simplest quantity to extract (by fitting to a decaying exponential),
there is the theoretical motivation that $m_f$, given by the position of a 
pole in the complex $k$-plane, is gauge-invariant, at least to all orders in
perturbation theory. As previously, we will
distinguish between propagation in temporal and spatial directions.

\subsection{Gauge Fixing}
\label{sec:gaugefix}

In order for the measurement of a gauge variant quantity such 
as the fermion propagator to be performed, we must impose a gauge condition
which selects a unique set of gauge configurations from the infinite number of
copies generated by local gauge transformations of the form
(in this subsection we will denote the lattice site by a suffix)
\begin{equation}
\theta_{\mu x}\mapsto\theta^\alpha_{\mu x}=\theta_{\mu x}+\partial_\mu\alpha_x,
\end{equation}
where on a lattice finite difference operators are defined:
\begin{equation}
\partial_\mu f_x=f_{x+\hat\mu}-f_x;\;\;\;
\bar\partial_\mu f_x=f_{x}-f_{x-\hat\mu};
\end{equation}
and $\alpha_x$ is any scalar function defined on the lattice sites.

For this study, we shall impose a latticised form of the Landau gauge condition
\begin{equation}
\sum_\mu \bar\partial_\mu\theta^\alpha_{\mu x} = 0,
\end{equation}
which is the extremum of
\begin{equation}
F^\alpha[\theta]=  \sum_{x} \sum_{\mu=1}^{3} (\theta_{\mu x}^{\alpha})^{2}, 
\label{eqn:numfunc}
\end{equation}
corresponding to the following functional in terms of continuum gauge fields: 
\begin{equation}
F[A]=\int d^3xA_{\mu}(x)A^{\mu}(x).
\end{equation}

In order to proceed, modifications will have to be made to this minimal gauge condition (henceforth referred to as mLandau gauge).  This is because it suffers from the so-called {\em Gribov ambiguity} \cite{Gribov:1977wm}. 

\subsubsection{The Gribov problem in QED$_3$}

When gauge fixing is performed non-perturbatively, it may always not be possible
to guarantee that there is a unique minimum of the functional $F[\theta]$. 
In numerical simulations, this can lead to a distortion of the results due to the underlying ambiguity \cite{Mitrjushkin:1996mq}.
The problem is normally associated with non-Abelian gauge fields in the continuum; however, it exists for Abelian fields on the lattice due to the toroidal boundary conditions \cite{Killingback:1984en}, 
which give rise to zero modes which cannot be removed by local gauge transformations 
and is especially acute for compact (cQED$_3$) formulations of the gauge fields as it allows for the existence of topological defects (such as double Dirac strings in $2+1$ dimensions or  double Dirac sheets in $3+1$ dimensions) 
whose creation or annihilation leaves the action unchanged \cite{Mitrjushkin:1996fw}.

Since we make use of a non-compact formulation of \qed (nc\qed$\!\!$) in this study, it seems that the only problem we might have to deal with is the former.  The {\em modified iterative Landau gauge} (miLandau gauge) \cite{Gockeler:1990bc,Durr:2002jc,Nakamura:1990dq} has often been used in order to deal with the problems due to the existence of zero-modes created by the boundary conditions of the lattice; however, it has not (as far as we are aware) been checked that there are any other sources of Gribov copies in this gauge. So, in what remains of this section, we shall describe miLandau gauge and present results that demonstrate that it does deal with the problem effectively, at least for the values of the parameters simulated in this paper.

\subsubsection{The miLandau gauge for ncQED$_3$}\label{sec:minLandau}

Firstly, we note that on the lattice we cannot rotate $\theta^{\alpha}_{\mu}
\mapsto \theta^{\alpha}_{\mu} + a_{\mu}$, where $a_{\mu}$ is an arbitrary
constant vector field, if we wish to preserve the gauge invariance of the
Polyakov and Wilson lines (defined to be products of the parallel transporters
$U_{\mu\,x}$ along contours which are closed by periodic boundary conditions in the temporal and spatial directions, respectively).  
Instead, the form of the allowed gauge rotations is restricted to $a_{\mu}=\frac{n 2 \pi}{L_{\mu}}$, where $n$ is an arbitrary integer.  

Using $\bar{\theta}_{\mu}=\frac{1}{V} \sum_{x} \theta_{\mu\,x}$ as the value of a constant background field (our zero-mode) we should expect the gauge degrees of freedom remaining after the mLandau gauge is fixed to vanish if we rotate 
\begin{equation}
\theta^{\alpha}_{\mu} \mapsto \theta^{\alpha}_{\mu} + \frac{n 2 \pi}{L_{\mu}}
\end{equation}
such that $-\frac{\pi}{L_{\mu}}<\bar{\theta}_{\mu}\le\frac{\pi}{L_{\mu}}$\footnote{A similar prescription, the Zero-Momentum Landau gauge \cite{Bogolubsky:1999cb}  sets $\bar{\theta}_{\mu}=0$.  The difference between this and miLandau gauge in the thermodynamic limit ($L_{\mu} \to \infty$) should be minimal \cite{Durr:2002jc}.}.

\begin{itemize}

\item  We fix mLandau gauge using a steepest descent algorithm \cite{Davies:1987vs}:
\begin{itemize}

	\item[$\diamond$] Given a gauge configuration \{$\theta$\},  for each
site we calculate the value of $G_x =\sum_{\mu} \bar{\partial}_{\mu}
\theta_{\mu x}$.

	\item[$\diamond$] If $\frac{1}{V} \sum_{x} G_x < R$, where $R$ the floating point value $10^{-6}$, we terminate the algorithm here. 
Otherwise, we continue.

	\item[$\diamond$] We rotate $\theta_{\mu x} \mapsto \theta_{\mu}(x) -
\partial_{\mu}\chi_x$ on every link of every lattice site, where $\chi_x=\eta G_x$, and  $\eta$ is a tunable parameter (here set to a value of $0.2$), used to optimise convergence.

	\item[$\diamond$] We repeat until the halting criterion is fulfilled.
\end{itemize}

\item  Once mLandau gauge fixing is complete, we calculate $\bar{\theta}_{\mu}$
for $\mu=\hat1$.

\item  If $\bar{\theta}_{\mu}\le -\frac{\pi}{L_{\mu}}$:
	\begin{itemize}
	\item[$\diamond$] Add $\frac{2\pi}{L_{\mu}}$  to each $\theta_{\mu x}$ until $-\frac{\pi}{L_{\mu}}<\bar{\theta}_{\mu}\le\frac{\pi}{L_{\mu}}$.
	\end{itemize}

\item	If $\bar{\theta}_{\mu}>\frac{\pi}{L_{\mu}}$:
	\begin{itemize}
	\item[$\diamond$] Subtract $\frac{2\pi}{L_{\mu}}$  to each $\theta_{\mu x}$ until $-\frac{\pi}{L_{\mu}}<\bar{\theta}_{\mu}\le\frac{\pi}{L_{\mu}}$.
	\end{itemize}

\item	Otherwise, leave each $\theta_{\mu x}$ unchanged.

\item Repeat the above for the remaining directions $\mu=\hat2,\,\hat3$.

\end{itemize}

\subsubsection{A test of this prescription in ncQED$_3$}

We wish to check that miLandau gauge removes Gribov copies from our 
measurements, by testing the effects of imposing miLandau gauge on randomly generated gauge copies of a set of gauge configurations \cite{Giusti:2001xf,Marinari:1991zv}.
The results were generated for $\kappa=1.00$ and $\kappa=10.00$ on a $16^{3}$ lattice for $\beta=0.2$ and $m=0.03$, the extreme values of the range at which we wish to measure the propagator.  200 mother configurations were generated, and for each mother we created three 500 configuration ensembles corresponding to one of the following random gauge transformations:
\begin{itemize}
\item[]
\begin{itemize}
\item[]
\begin{itemize}
\item[{\bf Group A:}] For all $x$ and $\mu$, $\theta_{\mu x} \mapsto \theta_{\mu
x} - \partial_\mu\alpha_x$ where $\alpha_x$ is a random number between $-$9 and 9.

\item[{\bf Group B:}] For all $x$ and $\mu$,  $\theta_{\mu x} \mapsto
\theta_{\mu x} + n_{\mu}\frac{2 \pi}{L_{\mu}}$, where $n_{\mu}$ is a random integer -- either 1, 0 or $-$1.

\item[{\bf Group C:}] We perform both the transformation performed on Group A and that performed on Group B.

\end{itemize}
\end{itemize}
\end{itemize}

\begin{figure}[htb]
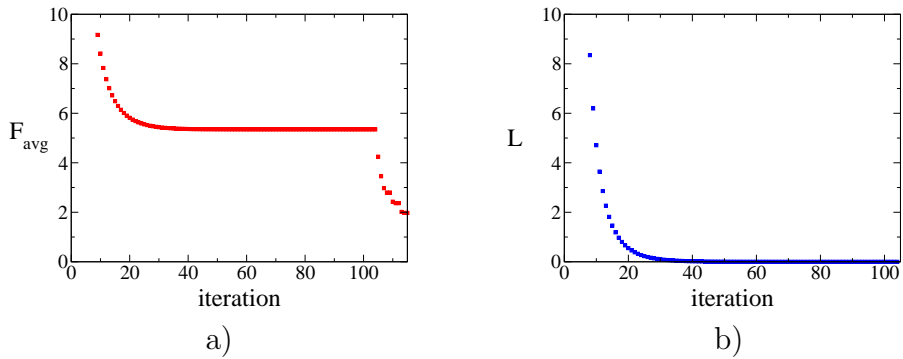

\vspace{.5cm}
\begin{center}
\epsfig{file=graphs/fthist.eps,height=4cm}
\hspace{1cm}
\epsfig{file=graphs/thetahist.eps, height=4cm}
\end{center}
\vspace{-.4cm}
\hspace{4.3cm}a)\hspace{6.4cm}b)
\vspace{-.1cm}
\caption{\small Typical behaviour of $F_{av}$ (a) and $L$ 
(b) during gauge fixing.  The plateau towards the middle of (a) corresponds to the area where (b) is converging on zero (that is to say, approaching mLandau gauge); the fall off beyond the hundred and fourth interaction is due to the imposition of full miLandau gauge.  The first few points of both plots have been omitted so that this behaviour is visible.}
\label{fig:typicalFavg}
\end{figure}

During the gauge fixing of each configuration, we monitored both the average value, $F_{av}$ of the gauge fixing functional (\ref{eqn:numfunc}) at each site
 and the value of the the function $L=\frac{1}{V} \sum_{x} G_x$, 
with $G_x$ 
defined in \S \ref{sec:minLandau}, for each iteration of the fixing.
The behaviour of these parameters for a typical configuration is shown in Figure \ref{fig:typicalFavg}.

We may also define a `variance' $dF$ \cite{Bogolubsky:1999cb}, which measures the difference in the minimised values of the gauge functional $F_{min}$ in a particular ensemble of a mother and associated daughter copies:
\begin{equation}
dF= \mbox{max}_{ij}[F_{min\: i} - F_{min\: j}].
\end{equation}
with $i, j =1,\ldots,N$ where $N$ is the number of daughter
configurations in the 
ensemble.
If there are no Gribov copies present, this quantity should be zero (more realistically, in a numerical simulation we expect it to be of the order of the residual, $10^{-7}$), otherwise we expect a large value.

\begin{figure}[htb]
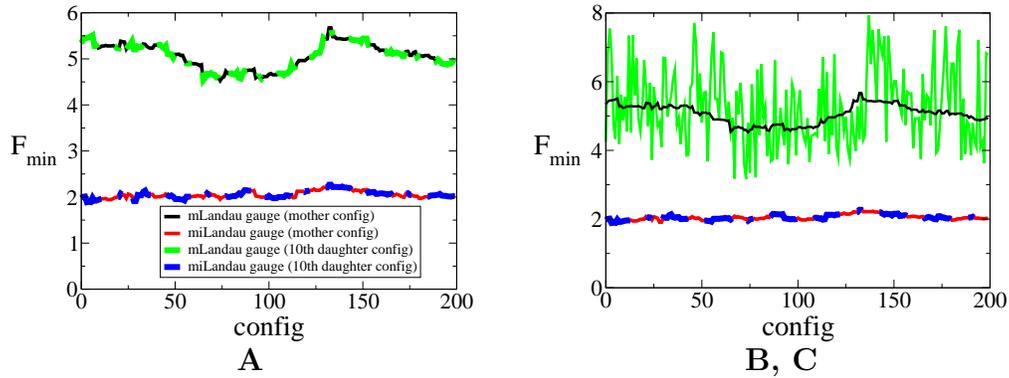

\vspace{1cm}
\begin{center}
\epsfig{file=graphs/norotatek1.eps, height=4.5cm}
\hspace{.5cm}
\epsfig{file=graphs/onlyrotatek1.eps, height=4.5cm}
\end{center}
\vspace{-.6cm}
\hspace{4.0cm}{\bf A}\hspace{6.4cm}{\bf B, C}
\vspace{.5cm}
\caption{\small Plots of $F_{min}$ at $\kappa=1.00$ for Groups A, B and C. $F_{min}$ for B and C are identical to within $10^{-7}$.}
\vspace{-.3cm}
\label{fig:Fmin_k=1.00}
\end{figure}

\begin{figure}[htb]
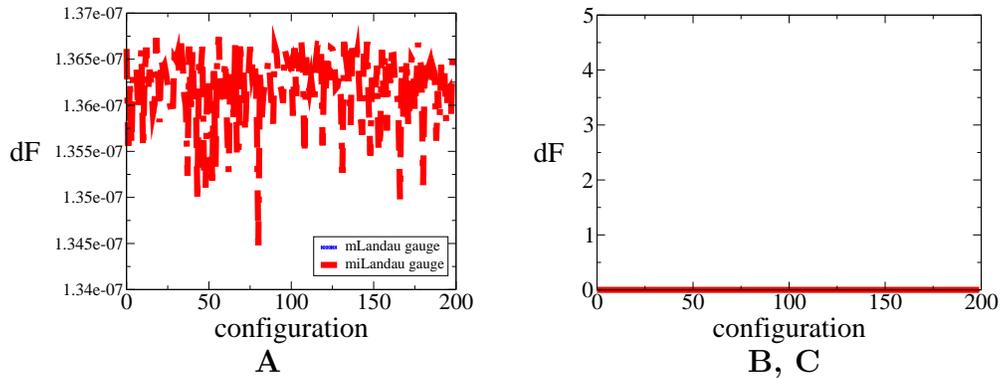

\vspace{1cm}
\begin{center}
\epsfig{file=graphs/norotvar_k1.eps, height=4.5cm}
\hspace{.5cm}
\epsfig{file=graphs/onlyvar_k1.eps, height=4.5cm}
\end{center}
\vspace{-.6cm}
\hspace{4.3cm}{\bf A}\hspace{6.2cm}{\bf B, C}
\vspace{.5cm}
\caption{\small Plots of the variance $dF$ at  
$\kappa=1.00$ for Groups A, B and C. $dF$ for B and C are identical to within $10^{-7}$. }
\label{fig:dFmin_k=1.00}
\end{figure}

The results of our simulations at $\kappa=1.00$ and $\kappa=10.00$ with respect
to Gribov copies were identical; we display figures for the former case, but our comments should be interpreted as generalising over both values of $\kappa$.
Figures \ref{fig:Fmin_k=1.00} and  \ref{fig:dFmin_k=1.00} plot $F_{min}$ and $dF$ for the ensembles generated using each
procedure.

\begin{itemize}
\item[]
\begin{itemize}
\item[]
\begin{itemize}
\item[{\bf Group A:}] Here we find that while the value of $F_{min}$ is
appreciably different in miLandau gauge from that in mLandau gauge (indicating
that zero modes exist and have been gauge rotated away in the former), $dF$ is of the order of $10^{-7}$, suggesting that the random gauge transformations used here do not usually generate Gribov copies.  

\item[{\bf Group B:}] Unlike the above case, here we can see that there are in fact Gribov copies in the mLandau gauge:  $dF$  is between 4 and 5 for $\kappa=1.00$ and 5 and 6 for $\kappa=10.00$.
However, this is not the case for miLandau gauge.
Here, as before, $dF$ in miLandau gauge is of the order of $10^{-7}$, and thus
we can conclude that it rids us of the Gribov copies introduced by
the random gauge transformation.

\item[{\bf Group C:}] Here, the crucial observation is that (to within $10^{-7}$), the results for these random gauge transformations are identical to those of Group B.  
Indeed, it would be worrying were it otherwise; the effects of the two sets of transformations should be additive, so one would expect only Group B's transformations to have any effect.
\end{itemize}
\end{itemize}
\end{itemize}

It is clear from the above that only shifts in the constant background field
appear to contribute to gauge copies, and these are readily dealt with through
the addition of further constraints to the minimal gauge fixing condition, via
the choice of miLandau gauge.
This stands in strong contrast to the case of cQED$_{3}$, where the compact Wilson gauge action allows for the existence of additional topological defects \cite{Mitrjushkin:1996fw} which are also solutions of the equations of motion and therefore are Gribov copies.

This `desert landscape' with respect to Gribov copies is not a disappointment --
in fact, it is precisely the situation desired; 
one can be sure that the gauge has been fixed as in an unambiguous
fashion.

\subsection{The Fermion Propagator}

In this section, we present measurements of the fermion propagator in the temporal and spatial directions:
\begin{eqnarray}
C_{f\mu}(x_\mu)&=&\sum_{\nu\not=\mu}\sum_{x_\nu=A}
\langle\chi(0)\bar\chi(x)\rangle,\nonumber\\
A&=& \left\{\begin{array}{cc} 
	2y_\nu& \mbox{Even numbered slice}\\
	2y_\nu + \hat{\mu} & \mbox{Odd numbered slice} \end{array}\right.
\label{eqn:fermcorr}
\end{eqnarray}
where the sum on $x$ only includes sites which are displaced from the origin by
an even number of lattice spacings in each of the two transverse directions
\cite{DelDebbio:1997dv,Gockeler:1991bu}. 
We have also imposed noncompact miLandau gauge using the procedure outlined in
\S\ref{sec:minLandau}.  As mentioned previously, the calculation of $C_{f\mu}$
required the generation of around 30,000 trajectories of mean length 1.0 per $\kappa$-point
in order to extract a signal from the considerable noise; for a dynamical
fermion simulation this amounts to a large effort, requiring between one and
three weeks per point to complete.  Because of this difficulty, the error bars
of our measurements remain sizable.

\subsubsection{Temporal propagator}

We extracted the fermion mass $m_{f\tau}$ 
in the temporal direction from the propagator data via the function
\begin{equation}
C_{f\, \tau}(\tau)=A(e^{-m_{f\, \tau}\tau}-(-1)^{\tau}e^{-m_{f\,
\tau}(L_\tau-\tau)})\label{eqn:tfermfit}
\end{equation}
using correlated least-squares fitting; the results are recorded in  Table \ref{tab:fermass} and Figure \ref{fig:tfermmass}.

\begin{figure}[htb]
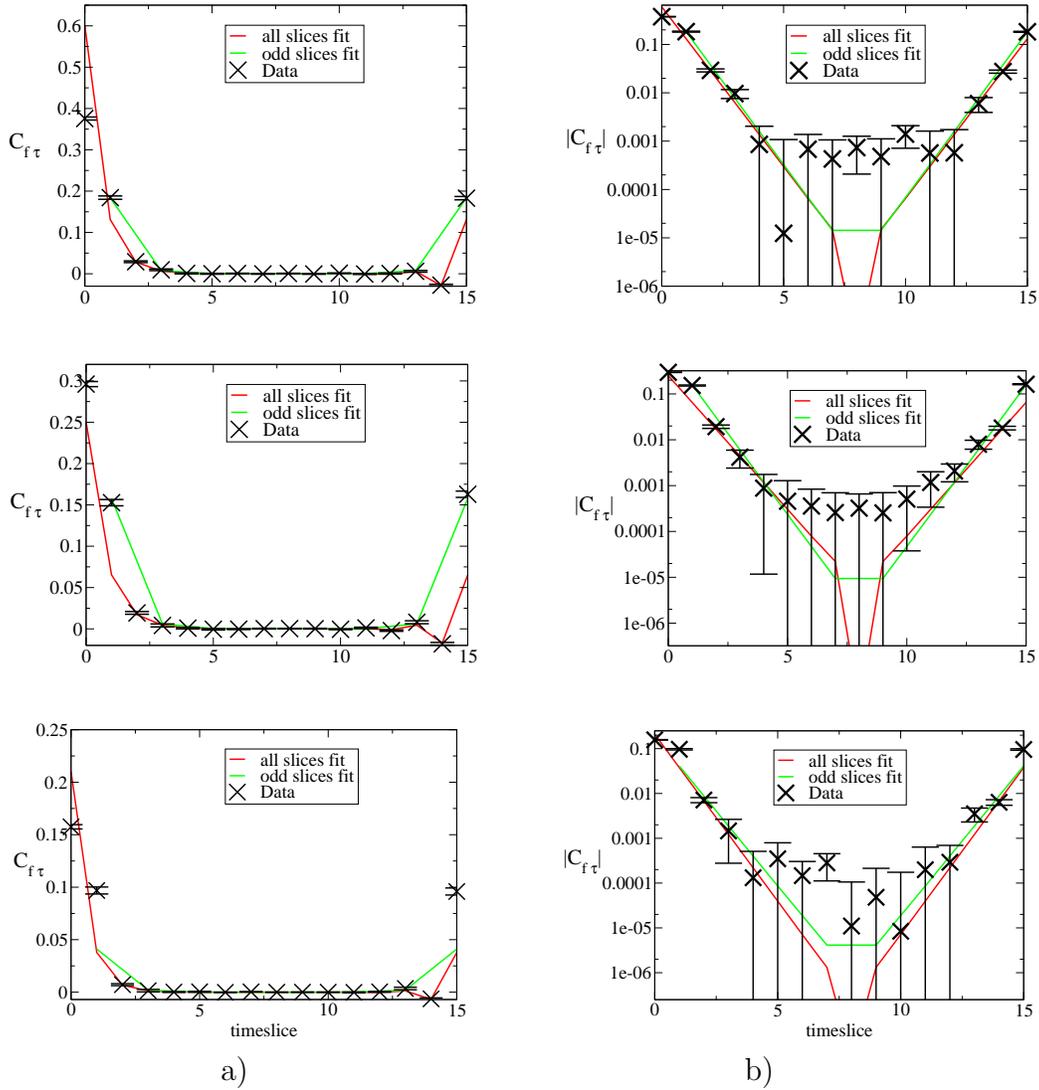

\vspace{1cm}
\begin{center}
\epsfig{file=graphs/tfermfitk6.eps, height=4cm}
\hspace{.9cm}
\epsfig{file=graphs/lntfermfitk6.eps,height=4cm}
\end{center}
\vspace{-.17cm}
\begin{center}
\epsfig{file=graphs/tfermfitk7.eps, height=4cm}
\hspace{1cm}
\epsfig{file=graphs/lntfermfitk7.eps,height=4cm}
\end{center}
\vspace{-.17cm}
\begin{center}
\epsfig{file=graphs/tfermfitk10.eps, height=4.2cm}
\hspace{1cm}
\epsfig{file=graphs/lntfermfitk10.eps,height=4.1cm}
\end{center}
\vspace{-.4cm}
\hspace{3.6cm}a)\hspace{6.6cm}b)
\caption{\small Comparison of all timeslice and odd timeslice fits on (a) a linear scale, and (b) logarithmic scales for $\kappa$ of 6.00, 7.00 and 10.00 in descending order  The errorbars represent unbinned, raw, statistical errors.}
\label{fig:fermfits}
\end{figure}

\begin{figure}[htb]
\vspace{0.5cm}
\begin{center}
\epsfig{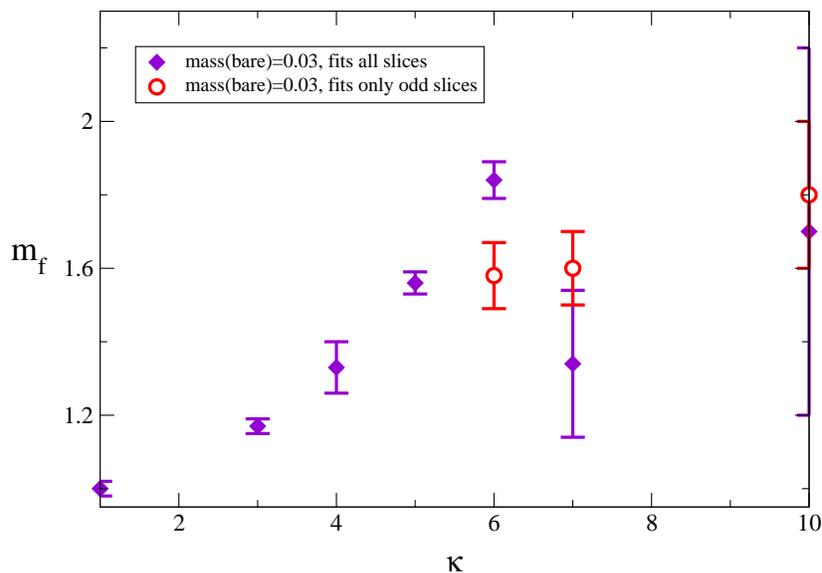}
\end{center}
\vspace{-.5cm}
\caption{\small The renormalised fermion mass, $m_f$, measured in Landau gauge on a $16^3$ lattice for $m=0.03$.}
\label{fig:tfermmass}
\end{figure}

First we should first examine Figure \ref{fig:fermfits}, which shows examples of
fermion propagators in the chirally restored phase $\kappa>0.5$. 
The following should be noted:

\begin{itemize}

\item  The central area of each propagator is fairly flat, with large error bars
(true of both phases). In this region the signal is overwhelmed by the noise and
is consistent with zero. The size of the window containing data points
exhibiting this behaviour decreased as the number of configurations in the
sample was increased, suggesting that the cause is insufficient statistics. 
Because of this, it proved necessary to use fitting windows that are wider than
the noisy region in order to extract a mass from the propagator. As in previous
studies of elementary fermion propagation \cite{DelDebbio:1997dv},
no indication of contamination 
from excited states was seen within those windows.

\item Figure \ref{fig:fermfits} also illustrates an interesting feature of the
fermion propagators for $\kappa>5$: the onset of a sawtooth-type behaviour
visible in the logarithmic plots that, although relatively small, grows more
pronounced with increasing $\kappa$.  Since it is hard to distinguish it from noise, we performed fits of (\ref{eqn:tfermfit}) to i) all of the timeslices and ii) to only the odd numbered timeslices for the propagators exhibiting this behaviour.
Ideally, a four-parameter fit is preferred to ii), but these proved to be unstable. 
\end{itemize}

The lines of best fit for both i) and ii) are included in Figure \ref{fig:fermfits} for purposes of comparison, and the masses extracted are included in Table \ref{tab:fermass} and Figure \ref{fig:tfermmass}.

It is worth discussing the origin of the sawtooth behaviour. The chiral
symmetry preserved by the lattice model
(\ref{eqn:lattact},\ref{eqn:fermion_matrix}) in the limit $m\to0$ is the U(1)
rotation
\begin{equation}
\chi(x)\mapsto\exp(i\beta\varepsilon(x))\chi(x);\;\;\;
\bar\chi(x)\mapsto\exp(i\beta\varepsilon(x))\bar\chi(x);
\end{equation}
where the phase $\varepsilon(x)\equiv(-1)^{x_1+x_2+x_3}$ distinguishes between
even ($e$) and odd ($o$) sites. In the chiral limit the 
only non-vanishing entries of the fermion propagator matrix are $M^{-1}_{oe}$
and $M^{-1}_{eo}$; for small but non-zero $m$ it should still be the case
in the chirally-symmetric phase 
that $\vert M^{-1}_{oe}\vert,\,\vert M^{-1}_{eo}\vert\gg\vert
M^{-1}_{ee}\vert,\,\vert M^{-1}_{oo}\vert$. In the timeslice correlator
$C_{f\mu}(x_\mu)$ defined in (\ref{eqn:fermcorr}) this implies that the signal 
should be much larger if $x_\mu$ is odd.
Figure \ref{fig:fermfits} shows that the sawtooth behaviour of the curve is not
especially pronounced, and it is at present unclear to what extent
the phenomenon is connected with the restoration of chiral symmetry.

\begin{table}[htb]
\centering
\setlength{\tabcolsep}{1pc}
\renewcommand{\arraystretch}{1.1}
\protect\footnotesize{
\begin{tabular}{|c|c|c|c|c|}
\hline
&$\kappa$ 	&$m$	&$\chi^2/d.o.f.$	& fit window\\
\hline
&1.00	&1.00(2)	&1.753		&2-14\\
&3.00	&1.17(2)	&1.696		&1-15\\
fit	&4.00	&1.33(7)	&0.919		&2-14\\
all timeslices	&5.00	&1.56(3)	&0.827		&1-15\\
&6.00	&1.5(2)		&1.211		&2-14\\
&7.00	&1.3(2)		&1.031 	&2-14\\
&10.00	&1.7(5)		&1.015		&2-14\\
\hline
\hline
fit only	&6.00	&1.58(9)	&0.546		&1-15\\
odd timeslices	&7.00	&1.6(1)		&1.228		&1-15\\
&10.00 	&1.8(2)		&0.863		&1-15\\
\hline
\end{tabular}}
\caption{\small Fermion masses $m_{f\,\tau}$ in the $\tau$ direction on a $16^3$ lattice, with $m=0.03$. }
\smallskip
\label{tab:fermass}
\end{table}

Fig.~\ref{fig:tfermmass} shows that for $\kappa \leq 5.00$, $m_{f \,\tau}$
increases with $\kappa$. The behaviour above $\kappa=5$ depends on the the type
of fit -- for fit ii) we see that the behaviour shows a non-zero mass in the
region, which is more or less constant . Fit i)
also shows the existence of a non-zero mass, but with more noise,
possibly since it does not account for the sawtooth behaviour.

Regardless of the method chosen for the fitting of the propagators, there is a
clearly a non-zero dynamically generated fermion mass in the chirally restored 
phase.  This is unexpected -- dynamical mass generation
usually implies $\chibchi\neq0$, and chiral symmetry restoration usually implies
massless fermions (Cf. Figs. 14 and 17 of Ref.~\cite{DelDebbio:1997dv}, 
illustrating light fermion propagation on a $16^3$ lattice 
in the chirally-symmetric phase of the 3$d$ Thirring
model). This seems to indicate that we are observing an unusual kind of chiral symmetry restoration; we shall return to this issue in due course.

\subsubsection{Spatial propagators}

We fit the spatial propagators to the following function: 
\begin{equation}
C_{f\mu}(x_\mu)=A(e^{-m_{f\, \mu}x_\mu}+(-1)^{x_\mu}e^{-m_{f\, \mu}(L_\mu-x_\mu)}):\label{eqn:spfermfitf}
\end{equation}
the change in sign compared to eqn.(\ref{eqn:tfermfit}) being due to the use of
periodic boundary conditions for the fermion fields in spatial directions,
and anti-periodic boundary conditions, consistent with the imaginary time
formalism used in (\ref{eq:BdG}), in the temporal direction.

\begin{figure}[htb]
\vspace{0.5cm}
\begin{center}
\epsfig{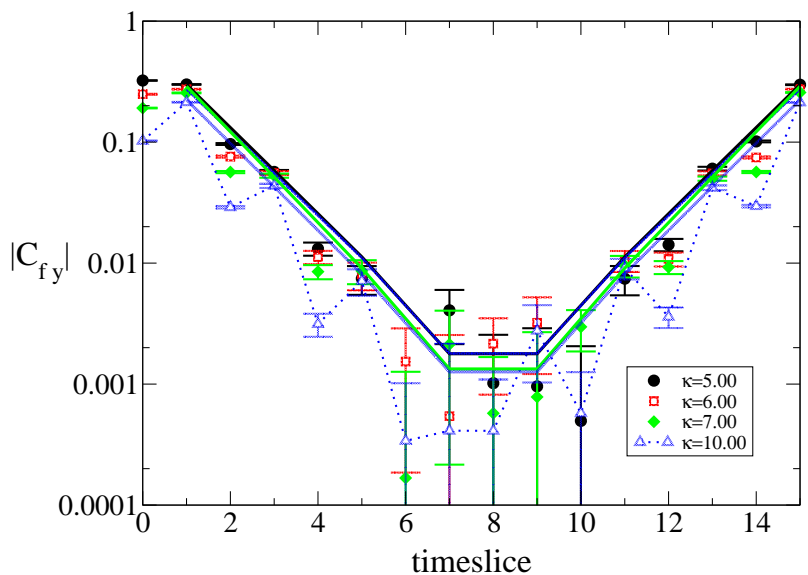}
\end{center}
\vspace{-.5cm}
\caption{\small Plot of the fermion space-slice propagator, for $5.00 \leq \kappa \leq 10.00$ on a $16^3$ lattice, along with the fitted curves.  Note how the sawtooth behaviour becomes more prominent as $\kappa$ increases, except on timeslices between 6 and 10, where noise dominates.  The errors here are unbinned.}
\label{fig:yfermsaw}
\end{figure}

Figure \ref{fig:yfermsaw} shows the absolute values of propagators in the
$y$-direction for $\kappa\geq 5.0$.  The propagators exhibit a more pronounced
form of the sawtooth behaviour than $C_{f\;\tau}$ ($C_{f\;x}$ do not exhibit 
this behaviour at all). Unlike in the case of the pions in \cite{Hands:2004ex},
this is not due to the fermion becoming light, as one can surmise from the slope of the curve.

\begin{table}[htb]
\centering
\setlength{\tabcolsep}{1pc}
\renewcommand{\arraystretch}{1.1}
\protect\footnotesize{
\begin{tabular}{|c|c|c|c|c|}
\hline
$\kappa$ 	&$m$	&$\chi^2/d.o.f.$	& fit window\\
\hline
1.00	&0.97(1)	&2.540		&1-15\\
3.00	&1.6(2)	&1.839	&2-14\\
4.00	&2.1(1)	&1.233		&1-15\\
5.00	&2.7(2)	&0.579		&1-15\\
6.00	&2.6(2) &1.097	&1-15\\
7.00	&3.6(7) &0.800	&1-15\\
10.00	&5(2)	&0.793	&1-15\\
\hline
\end{tabular}}
\caption{\small Fermion masses $m_{f\,x}$ in the $x$ direction on a $16^3$ lattice, with $m=0.03$. }
\smallskip
\label{tab:xfermass}
\end{table}

\begin{figure}[htb]
\vspace{0.5cm}
\begin{center}
\epsfig{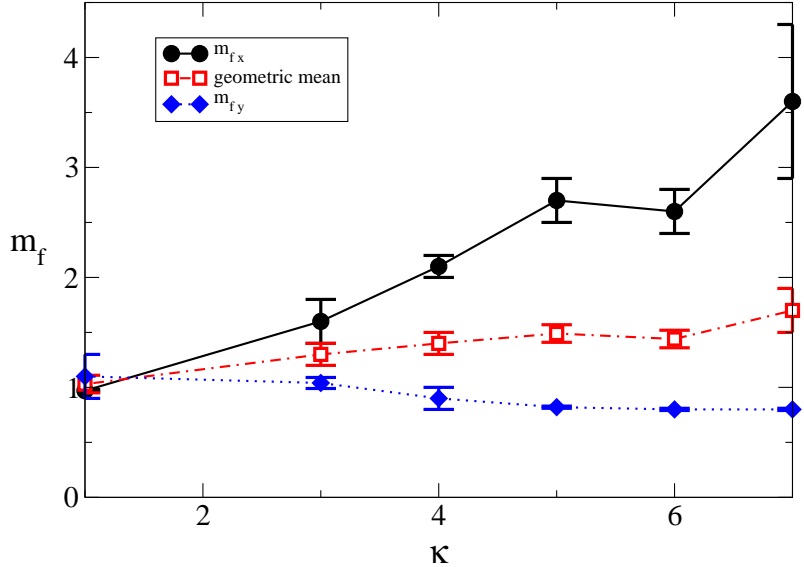}
\end{center}
\vspace{-.5cm}
\caption{\small Fermion screening masses $m_{f \,x}$, $m_{f\, y}$, 
and the geometric mean $\sqrt{m_{f\, x}m_{f\, y}}$, versus  $\kappa$, on a $16^3$ lattice.  Lines do not correspond to fits.}
\label{fig:alldirsferms}
\end{figure}
As for the sawtoothed  propagators in the $\tau$-direction, we performed fits
only to odd $y$, as four-parameter fits proved unstable.  The resulting
screening masses are shown in tables \ref{tab:xfermass} and \ref{tab:yfermass}.
\begin{table}[htb]
\centering
\setlength{\tabcolsep}{1pc}
\renewcommand{\arraystretch}{1.1}
\protect\footnotesize{
\begin{tabular}{|c|c|c|c|c|}
\hline
&$\kappa$ 	&$m$	&$\chi^2/d.o.f.$	& fit window\\
\hline
fit	&1.00	&1.1(2)	&0.887		&4-12\\
all timeslices	&3.00	&1.04(5)	&0.951		&3-13\\
	&4.00	&0.9(1)	&0.907		&4-12\\
\hline
\hline
&5.00	&0.82(1)	&2.475		&1-15\\
fit only	&6.00	&0.80(1)	&1.794		&1-15\\
odd timeslices	&7.00	&0.80(1)	&0.689		&1-15\\
&10.00 		&0.80(1)	&0.405		&1-15\\
\hline
\end{tabular}}
\caption{\small Fermion masses $m_{f\,y}$ in the $y$ direction on a $16^3$ lattice, with $m=0.03$.}
\smallskip
\label{tab:yfermass}
\end{table}
We can see that the fermions follow the same general trend
 as $\kappa$ increases as
the pions  in \cite{Hands:2004ex} -- those in the $x$-direction grow heavier,
and those in the $y$-direction grow lighter; anomalies in the data (e.g. at
$\kappa=6.00$) are likely to be due to noise. 

  The geometrical mean $\sqrt{m_{f\;x}m_{f\;y}}$
increases from 1 to $\sim1.75(20)$ as we move into regions of large anisotropy, which suggests
that some small dynamical effect may come into play over and above that of the
anisotropies themselves.  This could correspond to a renormalisation of the
parameter $\delta$ in Lee and Herbut's model to a value other than unity, as
$\sqrt{(\delta \lambda a)(\delta \lambda^{-1}a)}=\delta a$ \cite{Lee:2002qz}. 

\subsubsection{Renormalised anisotropy:}

\begin{figure}[htb]
\vspace{0.5cm}
\begin{center}
\epsfig{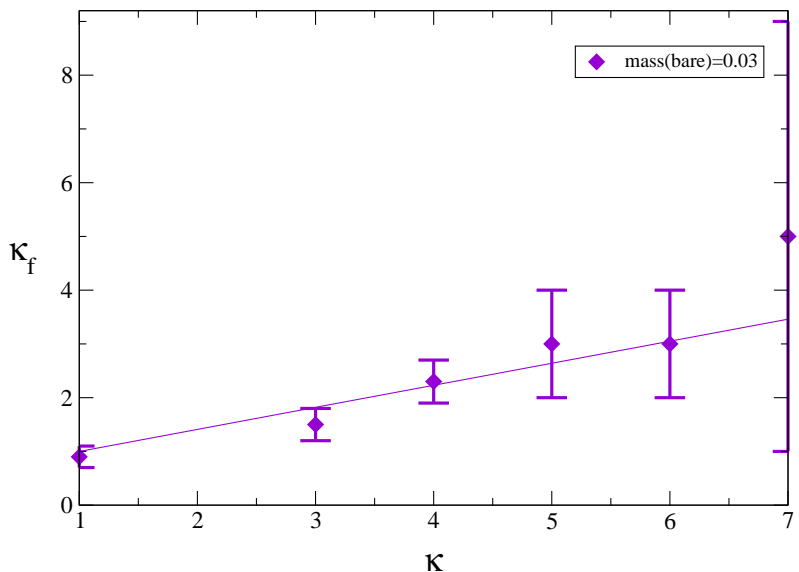}
\end{center}
\vspace{-.5cm}
\caption{\small The renormalised $\kappa$, $\kappa_{r\,f}$, together with a fitted curve, on a $16^3$ lattice.}
\label{fig:fermkrenormed}
\end{figure}

The renormalised fermion anisotropy $\kappa_{R\,f}= m_{f\,x}/m_{f\,y}$  is
displayed in Figure \ref{fig:fermkrenormed}.  It is a measurement of the
relevance of $\kappa>1$ relative to the particle in question.  We find that the
anisotropy parameter for fermions is {\em irrelevant} in the renormalisation
group sense; that is
\begin{equation}
R_{f}=\frac{(\kappa_{R\,f} -1 )}{(\kappa -1)}<1.
\end{equation}
Indeed, if we fit the above function to the data for $1\leq \kappa \leq 7.00$, we find $R_f=.41(4)$, with a $\chi^2/d.o.f.$ of 1.67.
This is striking, as the anisotropy is quite clearly relevant in the cases of
pions \cite{Hands:2004ex}, and scalars as shown in \S\ref{sec:krenscal}. 
The implication is that as $\kappa$ increases, the fermion -- anti-fermion bound
states become increasingly $1+1$-dimensional, only able to propagate in the $y -
\tau$ plane (in the original condensed matter-inspired model (\ref{eq:finalbit})
$f\bar f$ excitations associated with the other ``flavour'', ie. node pair,
would be confined to the $x -\tau$ plane).
 The only excitations able to explore the whole
$2+1$-dimensional space are the elementary fermions. This point will be further
discussed below.

\section{Discussion}
\label{sec:conclusion}

Here we summarise the main results of our study, and speculate as to the
behaviour
of QED$_3$ as the anisotropy $\kappa$ is increased.

We applied a finite volume scaling analysis to data from $16^3$, $20^3$ and
$24^3$ systems in an attempt to determine the order of the phase transition. 
There is no evidence for a diverging susceptibility as the volume increases, 
and remarkably the value 
$\kappa_c$ marking the apparent transition appears to be
very sensitive to system size; our fits assuming an isotropic model of finite
volume corrections yield $\kappa_c=7.66(5)$, which once the possibility of
anisotropic corrections is admitted drifts out to $\kappa_c=12.3(6)$. Since
the latter value lies outside our range of simulated parameters, it casts doubt
on our original claim \cite{Hands:2004ex} that a true 
chiral symmetry restoring transition is taking 
place. Rather, an interpretation of the transition in terms of a crossover from
strong to weak coupling regimes seems admissible -- very similar to the
transition observed in simulations of isotropic QED$_3$
\cite{Hands:2002dv,Hands:2004bh}. As in those studies, it appears to be a very
difficult task to determine computationally whether chiral symmetry is actually
broken in the weak coupling regime, reflecting the fact that QED$_3$ may be a
model with an abnormally large separation between the scale of dynamical
symmetry breaking $\Sigma$ and the natural mass scale $g^2$.
It should be stressed, however, that the studies of the pion and scalar spectra
in Sec.~\ref{sec:scalar}
are consistent with a chirally-restored vacuum at large $\kappa$.

What does seem clear is that any successful model of the finite volume scaling 
must take anisotropy into account -- here our analysis assumed weak
anisotropy, but models with differing critical exponents in different
directions cannot be excluded. Unfortunately, the cure for these many 
uncertainties is to accumulate data from many more values of $L$ and $m$, which
is beyond our current resources. 

However, it is intriguing to note that 
from \cite{Hands:2004ex,Sutherland}, we can estimate that at $T=0$ we enter the dSC phase (and \qed ceases to be a valid effective field theory of the cuprates) somewhere in the region  $6\lapprox\kappa\lapprox8$.  
Even taking
the isotropic estimate $\kappa_c=7.66(5)$ as the correct one, therefore,
it is uncertain whether the intermediate pseudogap phase between SDW and dSC can actually exist.
This raises a matter of some importance to future research: if $\kappa$ affects
the behaviour of the system\footnote{That is, if $\kappa$ is relevant, or if
$\kappa$ is irrelevant but $N_{fc}$ is not universal.}, 
does it do so {\em enough} to make a difference in the condensed matter systems for which anisotropic \qed is intended as an effective theory? 

Our studies of the propagation of $f\bar f$ bound states in the 
scalar channel showed evidence for degeneracy between scalar and pseudoscalar 
as $\kappa$ increases, although the propagator data are markedly noisier in the
scalar case. This is consistent with chiral symmetry restoration, but
bearing in mind the cautious note of the preceding paragraphs, 
we should note that a
very soft symmetry breaking cannot be excluded. Another important result is that
the renormalised anisotropy $\kappa_{R\sigma}\gapprox\kappa_{R\pi}$, implying
that anisotropy is a relevant perturbation for both sets of particles.
More graphically, this means that for large $\kappa$ $f\bar f$ bound states are
effectively constrained to propagate in just the $y$-direction, and their 
dynamics are essentially 1+1 dimensional.

The most significant result has emerged in the fermion sector, where we have
found evidence that dynamical mass generation persists even once the apparent
restoration of chiral symmetry has set in. Note that this result explains a
rather surprising result reported in \cite{Hands:2004ex}; namely, the
average plaquette action ${\beta\over2}\langle\Theta_{\mu\nu}^2\rangle$
increases with $\kappa$, implying that screening due to virtual $f\bar f$ pairs
in the quantum vacuum actually decreases with $\kappa$, in contradiction to what
would be expected if light fermion degrees of freedom were 
important in the high-$\kappa$ regime. The sawtooth structure that develops as
$\kappa$ increases may also be a sign of chiral symmetry restoration, although 
a study with $m$ varying, beyond our current resources, would be needed to
confirm this hypothesis.

The fact that a non-zero dynamically generated fermion mass accompanies the
chirally restored phase suggests that the symmetric phase is of an unusual
kind. Witten \cite{Witten:1978qu} has examined a similar situation in the Gross-Neveu model in $1+1$ dimensions;  $\chibchi=0$ and a dynamically generated mass may coexist if the following are the case:

\begin{itemize}
\item The physical fermion  is a branch-cut, not a pole, in momentum space, and lacks the same quantum numbers as the bare, massless fermion field $\psi$ (the former has zero chirality -- i.e. is chirally neutral -- whereas the bare fermion has a non-zero chirality).  It follows that chiral symmetry tells us nothing about the value of the dynamically generated fermion mass -- the system behaves in a chirally symmetric fashion in most respects apart from the existence of this mass.

\item There exists a massless (pseudo)scalar boson, which interacts strongly
with the fermion and carries the chiral current.  Interactions between it and
the $\psi$ field (which is distinct from the observed {\em physical} fermion)
are chirality changing.  The interaction between $\psi$ and the boson modifies
the chirally asymmetric portion of the fermion propagator, causing it to vanish.
\end{itemize}

It is important to note that the scalar field in this example is {\em not\/}
 a Goldstone boson, which must be weakly-interacting. 
In the $1+1$ dimensional 
Gross-Neveu model the formation of Goldstone bosons is prohibited by the
Coleman-Mermin-Wagner theorem \cite{Mermin:1966fe,Coleman:1973ci}, which states
the impossibility of spontaneously breaking a continuous global
symmetry in $1+1$ dimensions. 
A similar phenomenon has been observed in simulations of the 2+1$d$ Gross-Neveu
model at non-zero $T$ \cite{Hands:2001cs}.
While perhaps it not clear how to define the
effective
dimensionality of an anisotropic theory, we take from this analogy the
notion that infra-red fluctuations remain important in the chirally symmetric
phase; in other words the interaction between fermion and scalar degrees of
freedom  is strong.

In support of this hypothesis applying in the current situation, we point out
that the mass ratios $m_f:m_\pi:m_\sigma$ vary from $1:0.2:0.4$ at $\kappa=1$,
consistent with broken chiral symmetry, to $1.8:0.9:0.9$ at $\kappa=10$,
consistent with restored chiral symmetry, but in which the scalar bound states
are still tightly bound and light compared with the fermion mass scale. This
should be contrasted 
with the ``orthodox'' chiral symmetry restored scenario 
$m_f:m_\pi:m_\sigma\approx0.5:1:1$ 
observed in the 3$d$ Thirring model and portrayed in Fig.~17 of
\cite{DelDebbio:1997dv}.

It should be noted that the situation in which dynamical mass generation 
without symmetry breaking is observed is sometimes referred to as 
{\em pseudogap} behaviour\nolinebreak 
\footnote{We thank Kurt Langfeld for bringing this to our attention.}.  We must caution against confusing this with the pseudogap phase of the cuprate which we are modelling; while they share some behaviour in common (in both cases, we observe the phase disordering of an order parameter), they refer to different phenomena -- the former referring to a phase of the putative effective theory, and the latter to that of the behaviour of the full description of the superconductor from which it is derived. 
 
Another important observation in the fermion sector is that $\kappa_{Rf}<\kappa$
implying anisotropy is an irrelevant perturbation, i.e. fermions remain
2+1$d$ particles as $\kappa$ increases, although scalar-mediated interactions
among the fermions must be anisotropic. It will be a theoretical challenge to
formulate an effective description incorporating these features.

In many ways our study has raised more questions than it has answered; its
main results have not been predicted by analytic treatments of the system
performed so far.  This may raise questions regarding the conception of the
pseudogap in those models of HT$_c$ superconductivity -- such as that of
\cite{Franz:2002qy} --  which require the presence of massless fermions in the 
chirally symmetric phase, since it appears that the expected link between a
non-vanishing chiral condensate and a dynamically generated fermion mass 
is broken.
However, 
it is too early to make definitive statements; 
the fermion propagator should be measured in a number of gauges, so that we can be certain as to how much (if any) of the observed behaviour is an artefact of Landau gauge.
Ultimately, more data on how the dynamically generated fermion mass behaves as 
the chiral, thermodynamic and continuum limits are approached will be needed. 
Anisotropic QED$_3$ appears to be every bit as computationally demanding 
and as fascinating as its isotropic counterpart.

\end{document}